# Photo-Stimulated Electron Detrapping and the Two-State Model for Electron Transport in Nonpolar Liquids. [1]

Ilya A. Shkrob [a)] and Myran C. Sauer, Jr.

*Chemistry Division , Argonne National Laboratory, 9700 S. Cass Ave.,*

*Argonne, IL 60439*

**Abstract**

In common nonpolar liquids, such as saturated hydrocarbons, there is a dynamic equilibrium between trapped (localized) and quasifree (extended) states of the excess electron (the two-state model). Using time-resolved dc conductivity, the effect of 1064 nm laser photoexcitation of trapped electrons on the charge transport has been observed in liquid *n*-hexane and methylcyclohexane. The light promotes the electron from the trap into the conduction band of the liquid. From the analysis of the two-pulse, two-color photoconductivity data, the residence time of the electrons in traps has been estimated as ca. 8.3 ps for *n*-hexane and ca. 13 ps for methylcyclohexane (at 295 K). The rate of detrapping decreases at lower temperature with an activation energy of ca. 200 meV (280-320 K); the lifetime-mobility product for quasifree electrons scales linearly with the temperature. We suggest that the properties of trapped electrons in hydrocarbon liquids can be well accounted for using the simple spherical cavity model. The estimated localization time of the quasifree electron is 20-50 fs; both time estimates are in agreement with the "quasiballistic" model. This localization time is significantly lower than the value of 310±100 fs obtained using time-domain terahertz (THz) spectroscopy for the same system [E. Knoesel et al., J. Chem. Phys. **121**, 394 (2004)]. We suggest that the THz signal originates from the oscillations of electron bubbles rather than the free-electron plasma; vibrations of these bubbles may be responsible for the deviations from the Drude behavior observed below 0.4 THz. Various implications of these results are discussed.

---

[a] Author to whom correspondence should be addressed; electronic mail: shkrob@anl.gov.

I. INTRODUCTION

Detailed knowledge of the properties of excess electrons in dielectric solids and liquids is imperative for understanding the complex chemistry initiated by photo- or radiation induced ionization in such media. [1-7] In liquids comprised of nonpolar molecules with no electron affinity, such as saturated hydrocarbons, the excess electron is in a dynamic equilibrium: it neither dwells permanently at the mobility edge of the conduction band of the liquid (as a quasifree electron in simple liquids, such as Ar, Xe, and $CH_4$) [5,7] nor becomes permanently trapped or solvated, as occurs in polar liquids, such as water. [8] Instead, the electron constantly oscillates between the localized and extended states, accessing the latter by a thermal emission from shallow traps, in striking similarity to band-tail electrons in disordered semiconductors. In the absence of polar and/or readily polarizable functional groups, the attractive interaction of the electron with the solvent molecules is weak, and the binding energies $E_t$ of trapped electrons are just 50-250 meV, [1,2] as opposed to 1-2 eV in polar liquids and ammonia. [8-10] The thermal energy $k_B T$ at room temperature is ca. 26 meV; hence the ease of phonon-activated emission from the traps. This dynamic view of the electron is the essence of the two-state model for the electron transport in nonpolar liquids that emerged in the early 1970s. We address the reader to several reviews of the properties of the electron in such liquids, by Holroyd, [1,11,12] Schmidt, [2,4,7,13] and others, [3,6,14-17] which provide detailed discussion of this model and its variants. Below, we outline the main assumptions of the two-state model and key experimental observations supporting it.

The apparent electron drift mobility $\langle \mu_e \rangle$ in room temperature alkanes and cycloalkanes varies from $10^{-2}$ to 100 $cm^2/Vs$; the maximum mobility is observed for nearly spherical molecules, such as 2,2-dimethylpropane, the lowest - for long-chain normal alkanes, reaching $< 10^{-3}$ $cm^2/Vs$ for amorphous polyethylene. [2] There is a correlation between $\langle \mu_e \rangle$ and the activation energy for the mobility: the higher the energy, the lower is $\langle \mu_e \rangle$. [1] The latter can be estimated assuming that the quasifree electron, $e_f$, that exhibits high mobility ($\mu_f > 10$ $cm^2/Vs$) coexists with a trapped electron(s), $e_t$ (with a low mobility typical for molecular ions, $10^{-4}$-$10^{-3}$ cm$^2$/Vs). The model seeks to estimate the probability $P_f$ for the electron to remain in the quasifree state; the electron mobility is then given by $\langle \mu_e \rangle \approx \mu_f P_f$. Typically, a thermal equilibrium between the trapped and quasifree electrons is postulated. Let $\tau_f$ be the lifetime of the quasifree electron (controlled by the localization/trapping rate) and $\tau_t \gg \tau_f$ is the mean residence time in the trap; then, $P_f \approx \tau_f/\tau_t$ and $\langle \mu_e \rangle \approx \mu_f \tau_f/\tau_t$.[1,2] Assuming near constancy of $\mu_f$ and $\tau_f$ as a function of temperature, most of the temperature dependence for $\langle \mu_e \rangle$ is from that for the residence time $\tau_t$. The latter can be estimated from $\tau_t^{-1} = \nu \, \exp(-E_t/k_B T)$, where $\nu$ is the attempt-to-escape frequency ($10^{12}$-$10^{14}$ s$^{-1}$ [18-20] or $\nu \approx E_t/h$ [21,22]), i.e., the activation energy for the electron mobility, approximately equals $E_t$, the trap energy. The latter increases with $V_0$, the energy of quasifree electron at the mobility edge relative to the vacuum. The higher the $V_0$, the higher is the energy of detaching the electron from a solvation cavity into the conduction band.

For liquids in which the activation energy of electron transport is low and $\langle \mu_e \rangle >$ 1 cm$^2$/Vs, it is possible to determine the electron Hall mobility $\mu_H$. Unlike the drift mobility measured by d.c. [1,2,4-7,11-14] and a.c. (GHz) [3] conductivities, $\mu_H$ is the measure of the electron transport in the extended state only (the carrier has to move with a considerable velocity for the Lorentz force to bend its trajectory).[15] The typical Hall mobilities determined for liquids composed of nearly spherical alkane molecules are 10-100 cm$^2$/Vs.[15,23-25] Recently, mobilities of free *positrons* ($e^+$) in *n*-hexane and *iso*-octane (2,2,4-trimethylpentane) were determined from the Doppler shift of their 511 keV annihilation line,[26] yielding ca. 50 and 70 cm$^2$/Vs, respectively (the mobilities of free electrons and positrons must be similar). For *iso*-octane, $\mu_H$ is ca. 20 cm$^2$/Vs at 270-400 K, which is ca. 3 times the apparent drift mobility $\langle \mu_e \rangle$.[25] There are alkane liquids (e.g., neopentane and 2,2,4,4-tetramethylpentane) for which the two mobilities are virtually the same.[15,25] It is generally believed that the mobility of quasifree electrons in saturated hydrocarbons should be of the same order of magnitude as $\mu_H$, viz. a few tens of cm$^2$/Vs. Similar estimates (30-400 cm$^2$/Vs) were obtained theoretically, e.g. by means of density

4.

fluctuation model based on the Cohen-Lekner scattering theory, [27] by Berlin et al. [28] More recently, Mozumder [21,22,29,30] pointed out that in many systems, electron trapping occurs *faster* than the relaxation of the electron velocity in the extended state, thereby dramatically reducing the apparent mobility in the free state from its maximally attainable value of ca. 100 cm$^2$/Vs (the "quasiballistic" model). For *n*-hexane, this model predicts $\tau_f \approx 30$ fs and $\tau_t \approx 11$ ps. [29]

Importantly, there are variants of the two-state model that postulate no time scales for the equilibration between the trapped and free electrons. Schiller [31] and others [28,32,33] suggested that the energies of localized states follow a normal distribution around their mean value with the dispersion $\sigma$ of ca. 120 meV at 295 K (see Appendix B in the Supplement [34] and refs. 28, 31, 32 and 33). Since only the states with positive binding energy are stable towards delocalization, $P_f \approx erfc(E_t/\sqrt{2}\sigma)$. [31] This simple approach accounts reasonably well for the temperature [28] and pressure [33] dependencies of $\langle \mu_e \rangle$, without invoking explicitly the electron equilibria, provided that the changes in the energy $V_0$ and the free-electron mobility $\mu_F$ are taken into account.

While the two-state model captures the most important facet of electron transport in nonpolar liquids (the coexistence of extended and localized states), it has several weaknesses. First, the postulated equilibrium of free and trapped electrons has not been observed experimentally on its assumed time scale. The equilibrium is deduced solely from the temperature dependencies of average electron properties, such as $\langle \mu_e \rangle$ and the rate constants for electron attachment to solute molecules. As far as the latter properties are concerned, it is possible to formulate a self-consistent two-state model without invoking any *electron* equilibria. [28,31-33] Second, the model does not specify the nature of electron traps, such as their structure, availability, and energetics; it also does not address the possibility of trap-to-trap migration. [1,2] Third, it does not explicitly address the relaxation of the solvent (which is assumed to be very rapid) and concurrent changes in the transport properties of the electron during its localization and trapping.

These recognized deficiencies of the two-state model have been brought to a sharp focus by the recent time-domain terahertz (THz) spectroscopy experiments of Knoesel et al. [35,36] The authors have interpreted their results as the first *direct* observation of a quasifree electron in liquid alkanes. The THz spectra were accounted for by a Drude model for a dilute free-electron plasma with a damping (scattering) time $\tau_d \approx 310 \pm 100$ fs (for *n*-hexane and cyclohexane). Obviously, this scattering time cannot be longer than the trapping time $\tau_f$. In the Drude model of an electron plasma, the mobility $\mu_f$ of the quasifree electron is related to the scattering time as $\mu_f = e\tau_d/m_e^*$, where $e$ is the elementary charge and $m_e^*$ is the effective electron mass. Assuming that the latter equals the electron rest mass $m_e$, one obtains an estimate of $\mu_f \approx 560$ cm$^2$/Vs. [36] This estimate is more than an order of magnitude greater than the estimates obtained from the Hall [15,23] and positron [26] mobility measurements. If one equates the scattering time $\tau_d$ and the trapping time $\tau_f$ (as suggested by Knoesel et al.) [35,36] and takes into account that $\langle \mu_e \rangle \approx (7-9) \times 10^{-2}$ cm$^2$/Vs at 295 K, [3,37-39] an estimate of $\tau_t \approx 1.7$ ns is obtained using the equilibrium two-state model. Thus, it appears that the trapping time is by an order of magnitude and the detrapping time is by two orders of magnitude greater than predicted by Mozumder using "quasiballistic" model. [22,29] Furthermore, assuming that the quasifree electron has thermal velocity, the free path $\Lambda$ of this electron may be estimated from $\mu_f \approx \left(8e^2/9\pi m_e^* k_B T\right)^{1/2} \Lambda$, [28] which yields $\Lambda \approx 400$ Å. The latter estimate is comparable with the Onsager radius $r_c = e^2/\varepsilon k_B T$ [1,3] for electron-hole pairs in the *n*-hexane (ca. 300 Å for the static dielectric constant $\varepsilon \approx 1.8$). Thus, it appears that electrons generated by the photoionization of *n*-hexane would typically avoid the recombination with their parent holes, in stark contradiction of experimental observations, which suggest that >95% of the ion pairs recombine geminately. [1,3,36] All of the above illustrates the problem that the THz observations of Knoesel et al. [35,36] present for the current picture of electron dynamics in saturated hydrocarbons. The root cause of this problem is that the scattering time (localization time?) for the free electron turned out to be unexpectedly long.

In this work, we attack the problem from the opposite end: specifically, we estimate the residence time $\tau_t$ of the electrons trapped in neat *n*-hexane and methylcyclohexane. Knowing these estimates and the equilibrium mobility, it is possible to estimate the free electron lifetime $\tau_f$. Our estimates for $\tau_t$ (ca. 10 ps at 295 K) appear to be in good agreement with the detrapping times predicted by Mozumder;[29] on the other hand, these estimates are inconsistent with the estimates obtained from the results of Knoesel et al.[35,36] In the companion paper, deeper (400-800 meV) traps for electrons were introduced by addition of polar molecules, and the residence times and energetics for these traps determined.[40] Once more, simple estimates for detrapping rate worked well for these modified traps. We demonstrate that the energetics of electron traps, as determined from the known optical spectra of electrons in alkane liquids using the spherical well ("electron bubble") model discussed in Sec. V.A correlates very well with the activation energies of electron transport. The trapped electrons involved in the conduction in liquid alkanes are the same as the electrons observed spectroscopically. There appear to be no inconsistencies regarding the energetics and properties of such trapped electrons. We, therefore, suggest that the THz signal observed by Knoesel et al.[36] does not originate from quasifree electrons. Rather, it originates from the oscillations of the electron bubbles in the THz electric field (Sec. V.C). Simple estimates are given to corroborate this reinterpretation of the THz data. To reduce the length of the paper, some figures and sections are placed in the Supplement. The figures with the designator "S" (e.g., Figure 1S) are placed therein.

**II. THE CONCEPT OF THE EXPERIMENT.**

Consider a trapped electron in equilibrium with the quasifree electron. We will assume that $\tau_f \ll \tau_t$ (i.e., the equilibrium concentration of *quasifree* electrons $e_f^0 \approx (\tau_f/\tau_t) e_0 \ll e_0$, the total molar concentration of the electrons), so that the mean conductivity signal $\kappa_0 = F\mu_f e_f^0$ is given by

$$\kappa_0 = F\langle\mu_e\rangle e_0 = F\mu_f \tau_f e_0/\tau_t, \qquad (1)$$

where $F$ is the Faraday constant. We will assume that a sufficiently long laser pulse (with pulse irradiance $J(t)$ and pulse duration $\gg \tau_p$) photoexcites the trapped electron and promotes it into the conduction band of the liquid. The resulting "hot" electron with

7.

mobility-lifetime product $\mu_h \tau_h \ll \mu_f \tau_f$ rapidly thermalizes yielding a quasifree electron. Let $\sigma$ be the cross section for the conversion of the trapped electron into the quasifree electron. During the photoexcitation, the electron concentrations change as

$$de_{f,t}/dt = \pm \sigma J(t) e_t \pm \left( e_t/\tau_t - e_f/\tau_f \right). \tag{2}$$

For a long excitation pulse, we may assume quasi-stationary conditions, so that $e_t(t) \approx e_0$, $e_f(t) = \tau_f \left[ \tau_t^{-1} + \sigma J(t) \right] e_0 \ll e_0$ and the change in the conductivity signal

$$\Delta \kappa(t) = F \mu_f \left( e_f(t) - e_f^0 \right) = F \mu_f \tau_f \sigma e_0 J(t). \tag{3}$$

The total integral $\Delta A$ under the photoinduced signal is given by

$$\Delta A = \int dt \ \Delta \kappa(t) = F \left( \mu_f \tau_f \right) \sigma J e_0 \tag{4}$$

where $J = \int dt \ J(t)$ is total photon fluence of the laser pulse. The ratio $r$ of the integral $\Delta A$ to the equilibrium conductivity signal is given by

$$r = \Delta A / \kappa_0 = \sigma \tau_t J, \tag{5}$$

Therefore, if the cross section $\sigma$ is known, $\tau_t$ can be determined directly from the ratio $r$. Note that the knowledge of the parameters $\mu_f$ and $\tau_f$ and the electron concentration is not needed to determine $\tau_t$. Furthermore, once the detrapping time $\tau_t$ is known, the product $\mu_f \tau_f$ and the mean square free path

$$\Lambda_t = \left( 6 \mu_f \tau_f k_B T / e \right)^{1/2} \tag{6}$$

of the quasifree electron (in the diffusive model) can be estimated. As shown in Appendix A in the Supplement, eq. (5) holds under more general assumptions than given above. In particular, it can be shown that for a system with many electron traps with different cross sections $\sigma$ and residence times $\tau_t$, $r/J \approx \langle \sigma \tau_t \rangle$, where $\langle ... \rangle$ stands for

8.

averaging over all electron populations at equilibrium. Using the same notation, the apparent electron mobility $\langle \mu_e \rangle \approx \mu_f \tau_f \langle \tau_t^{-1} \rangle$.

Eqs. (4) and (5) have already been derived, albeit in a different form, by Yakovlev and co-workers, [6,20,41] who studied photoinduced electron detachment from low-temperature hydrocarbons; the study of room temperature liquids was prohibited by inadequate time resolution of their setup. Balakin et al. [20] used 694 nm photoexcitation of electrons in *iso*-octane at $T$=175 K ($\langle \mu_e \rangle \approx 1.7$ cm$^2$/Vs) to estimate $\tau_t \approx 22\pm6$ ps, $\mu_f \tau_f \approx (3.8\pm1)\times10^{-11}$ cm$^2$/V, $\Lambda_t \approx 180\pm30$ Å, and $\nu \approx 3.5$ ps$^{-1}$ for the attempt-to-escape frequency from $E_t \approx 60$ meV [2] traps. For 1060 nm photoexcitation of *n*-hexane and methylcyclohexane at 180-185 K (at which $\langle \mu_e \rangle$ is ca. $10^{-3}$ and $5.5\times10^{-4}$ cm$^2$/Vs, respectively), Balakin and Yakovlev [41] obtained $\mu_f \tau_f \approx (4\pm2)\times10^{-13}$ and $4.7\times10^{-13}$ cm$^2$/V, respectively, which corresponds to $\tau_t$ of 400±200 and 850 ps and $\Lambda_t$ of 16 and 21 Å, respectively. The short free path in low-mobility alkanes suggests that $\mu_f < 30$ cm$^2$/Vs, which Yakovlev and Lukin [41] found unreasonably low, though such estimates are in good agreement with the subsequently determined $e^-$ Hall [15,23-25] and $e^+$ drift [26] mobilities. The smallness of the product $\mu_f \tau_f$ is readily accounted for in the "quasiballistic" model of electron transport [22] as an indication of the regime in which the relaxation of the velocity of free electron takes longer than its localization. Yakovlev and Lukin, [6] however, suggested that the product $\mu_f \tau_f$ in eq. (4) should be replaced by a sum $(\mu_h \tau_h + \mu_f \tau_f)$ of the mobility-lifetime products for the "hot" and quasifree electron, respectively. They speculated that the first term in the latter expression makes the largest contribution to $\Delta A$. As stated above, this assumption is not required in the light of the later findings; however, even if that were the case, the method would still yield an estimate for $\tau_t$ from the *above*. The greater problem with these previous measurements is that the cross sections $\sigma$ for the electron photodetachment were estimated from the spectra of radiolytically generated electrons in room-temperature *n*-hexane and methylcyclohexane (assuming unity quantum yield for the bound-to-continuum transition). A subsequent study by Atherton et al. [42] indicated that the absorptivity of the electron in methylcyclohexane at 1 μm increases as the temperature decreases, and this

9.

trend (concurrent with the blue shift of the absorption spectrum) [42] agrees well with the theoretical analyses in the "bubble model" [9,16,43-47] of solvated electron discussed in Sec. V.A. This arbitrariness decreases the confidence in the estimates of Yakovlev and co-workers. [6,20,41] Thus, we aimed to explore the region near room temperature, for which better estimates of all the parameters involved are available. The additional benefit of studying this region is the ability to directly compare our estimates with those obtained theoretically [21,22,29] and from the THz experiments. [35,36] To this end, improvements in the time resolution and the sensitivity of conductivity measurements were required.

### III. EXPERIMENTAL.

*n*-Hexane and methylcyclohexane (99+%, Aldrich), and *iso*-octane (Baker) were passed through activated silica gel to remove olefin impurity and, for methylcyclohexane, 0.05 vol % of toluene. From gas chromatography, after the silica gel treatment, *n*-hexane still contained traces of *n*-pentane, 3-methylpentane, methylcyclopentane, and 2,2-dimethylpentane (<0.2 vol% in total). The methylcyclohexane contained traces of dimethylcyclohexanes and polymethylated cyclopentanes. By deliberately adding these chemicals, we found that these impurities had no effect on our conductivity measurement. The probable reason is that all impurity alkanes exhibit similar mobility and activation (trap) energy for electron migration to the main component. The measurements of the electron mobility were carried out in $N_2$- or Ar- saturated solutions.

The conductivity setup was similar to that described in our previous publications. [48-50] Fifteen nanoseconds fwhm pulses of 248 nm photons from a Lambda Physik LPX 120i laser were used to ionize either neat alkanes or 5 μm anthracene, via their biphotonic excitation. The neat hydrocarbon liquids were photolyzed in a cell with 4 cm optical path, and the anthracene solutions (used only to obtain the temperature dependence) were photolyzed in a 2 cm path cell. Both cells have two planar *Pt* electrodes spaced by 0.65 cm to which a constant voltage of 4-5 kV is applied. The 2 cm cell was placed in an aluminum jacket; the temperature of the sample was regulated by circulating water through this jacket. The entire setup was put in an aluminum box purged by dry air. The collimated 248 nm beam entered the cell from one end through a 3 mm diameter aperture; a collinear beam of 532 nm or 1064 nm photons from a Quantel Brilliant Nd:YAG laser (6 ns fwhm pulse with a Gaussian time profile) entered the cell from the opposite end through a 4 mm diameter aperture. The 1064 nm (or 532 nm) beam completely enveloped 248 nm beam inside the conductivity cell. Appropriate optics were used to maintain the collimation and collinearity along the optical path of the two beams inside the cell. The conductivity signal was terminated into 50 Ω, amplified by 20-30 dB and recorded using a Tektonix DSA-601 digitizer. The time resolution was better than 2 ns. The delay time $t_L$ of the 1064 nm pulse relative to the 248 nm excitation pulse was 25-800 ns; the time jitter between the 248 and 1064 nm pulses was < 3 ns. The acquisition electronics was typically triggered by the 1064 nm pulse.

The maximum pulse energy of the 1064 nm light transmitted through the 4 mm aperture was 190 mJ, and the maximum photon fluence $J$ through this aperture was 1.5 J/cm$^2$ (or 9x10$^{18}$ photons/cm$^2$). This large fluence was needed in order to observe the $\Delta\kappa(t)$ signal due to the smallness of the parameters $\tau_t$ and $\sigma$ for trapped electrons near 295 K (see below). The photon fluence of 248 nm light was < 0.1 J/cm$^2$, and the typical electron concentration was 5-10 nM (in neat *n*-hexane). The lifetime of the electron (typically 300-500 ns) was controlled by an electron-scavenging (e.g., oxygen) impurity (in the anthracene solution, electron attachment to the aromatic photosensitizer also contributed to limit the lifetime). A typical rate constant for such a reaction in *n*-hexane is (1-2)x10$^{12}$ M$^{-1}$ s$^{-1}$.[3] Cross recombination in the bulk and the movement to the electrodes of electrons and ions were negligible for $t <$ 1 µs under our excitation conditions. Geminate recombination for electron-hole pairs in *n*-hexane and methylcyclohexane is complete well within the duration of the 248 nm pulse (the Onsager times [of diffusional travel over a distance $\approx r_c$] for these two liquids at 295 K are ca. 4 ns and ca. 50 ps for *n*-hexane and *iso*-octane, respectively).

To determine the $\Delta\kappa(t)$ signal, the 1064 nm (or 532 nm) laser was pulsed on and off while the 248 nm laser was pulsed for every shot, and the corresponding signals $\kappa_{on}(t)$ and $\kappa(t)$ were subtracted. A small signal (contributing < 1 % to this difference signal) induced by the action of the 1064 nm laser alone, through the pickup of radio frequency noise from the laser Q-switch, was subtracted from the $\Delta\kappa(t)$ signal.

Continuous laser photolysis causes the accumulation of photoactive products in the solution; care was exercised to minimize their interference by frequent replacement of the sample. If not specified otherwise, the measurements were carried out at 295 K. The conductivity is given in units of nS/cm (= $10^{-7}$ $\Omega^{-1}$ $m^{-1}$).

**IV. RESULTS.**

A typical conductivity signal from neat *n*-hexane following 248 nm biphotonic laser excitation of this solvent is shown in Fig. 1 (to the left). Over the first 1 µs after the ionization event, the electron is scavenged by an impurity in the solution. This decay is single exponential (Fig. 2), and the conductivity signal $\kappa_{ion}$ at the later delay times is from the secondary ions (this signal has been subtracted from $\kappa(t)$ in Figs. 1 and 2). For $5\times10^{-4}$ M triethylamine in *n*-hexane saturated with $SF_6$ (used to scavenge the electrons, converting them to $F^-$), our time-of-flight measurement yielded $9\times10^{-4}$ and $6\times10^{-4}$ cm$^2$/Vs for the mobility of the positive (triethylamine$^+$) and negative ($F^-$) ions, respectively. Thus, the combined ion mobility $\mu_i$ was ca. $1.5\times10^{-3}$ cm$^2$/Vs. Similar estimates of ion mobility in *n*-hexane were reported by others (e.g., see Table 12.3 in ref. 51 and refs. 52). The electron signal is ca. 56 times greater than the signal from the ions, suggesting that $\langle\mu_e\rangle \approx 0.085$ cm$^2$/Vs. (vs. 0.082 cm$^2$/Vs obtained in ref. 37). Direct time-of-flight estimates for the electron mobility in high-purity *n*-hexane at 295 K are between 0.073 and 0.092 cm$^2$/Vs. [3,37-39]

Balakin et al. [20] used 694 nm (1.79 eV) light to excite trapped electrons in cold *iso*-octane and some of our initial experiments with saturated hydrocarbons were carried out using 532 nm (2.33 eV) light (Figure 1S). This turned out to be problematic since the 532 nm light excited one of the impurity anions, perhaps $O_2^-$ (see Sec. 1S of ref. 48). The latter anion is known to absorb across the entire visible: the photodetachment threshold is ca. 2 eV and the cross section near this threshold is ca. $(1-10)\times10^{-19}$ cm$^2$. [53,54] Since the thermal emission of the electron from $O_2^-$ is very slow, [54] the 532 nm photodetachment results in a sudden stepwise increase in the conductivity signal; the photoinduced signal $\Delta\kappa(t)$ decays in exactly the same way as the conductivity signal $\kappa(t)-\kappa_{ion}$ itself, save

12.

for the delay time. The initial $\Delta\kappa(t \approx t_L)$ signal plotted against the delay time $t_L$ of the 532 nm pulse mirrors the decay kinetics $\kappa(t) - \kappa_{ion}$ of the electron, since the anion is generated when the electron is scavenged by an impurity (ref. 48 and Figure 1S). Intentional addition of traces of oxygen increases the photoinduced signal. Due to the strong interference from this $\Delta\kappa(t)$ signal, only 1064 nm photoexcitation can be used for quantitative measurements.

Figures 1 and 2 demonstrate a typical $\Delta\kappa(t)$ signal induced by the absorption of 1064 nm light in room-temperature photoionized *n*-hexane. This signal consists of two components: (i) a fast component whose time profile follows that of the 1064 nm laser pulse (corrected by the response function of the setup) and (ii) a slow component whose long-term decay kinetics are identical to those for the conductivity signal $\kappa(t) - \kappa_{ion}$ itself (this slow signal is very small for $t_L = 45$ ns in Fig. 2). The overall kinetics of $\Delta\kappa(t)$ is given by a sum of a Gaussian "spike" (for the fast component) and the same Gaussian convoluted with an exponential (for the slow component); at any delay time $t_L$ of the 1064 nm pulse, these two signals can be to separated using least squares fitting, as shown in Fig. 2S, part (a). The weight of the fast component (empty circles in Fig. 1), decreases with increasing delay time $t_L$ in direct proportion to $\kappa(t) - \kappa_{ion}$, whereas the weight of the slow component *increases* in the same direction, eventually saturating at delay times when all electrons are scavenged (see Figs. 2 and 2S(a) for *n*-hexane and Fig. 2S(b) for methylcyclohexane). This behavior suggests that the slow component is due to the photodetachment from an impurity anion. Balakin and Yakovlev [41] observed the same two components at 180-185 K and interpreted the fast component as the signal from the electrons photodetached from intrinsic, shallow traps and the slow component as that from extrinsic, deep traps. Since the lifetime $\tau_f$ of the quasifree electron is much shorter than the duration of the pulse and all electron equilibria settle well within 1 ns, the fast component follows the time profile of the 1064 nm pulse.

The power dependencies for the photoinduced signals also suggest different origins for these two components. To obtain the power dependence for the slow component, the $\Delta\kappa(t)$ signal attained at the end of the 1064 nm laser pulse for $t_L = 760$ ns (extrapolated from the exponential "tail" as shown in Fig. 3S(a)) was plotted against the average fluence $J$ of the 1064 nm photons (see Fig. 3S(b)). Due to the reaction with an impurity (with time constant of ca. 270 ns), 94% of the electrons were converted to anions at this relatively long delay time. The plot $\Delta\kappa(t \approx t_L)$ vs. $J$ is exponential, $\Delta\kappa \approx \Delta\kappa_\infty (1 - \exp[-\sigma_i J])$, as would be expected for the photodetachment from a molecular anion with $\sigma_i \approx (4.6\pm0.4) \times 10^{-19}$ cm$^2$. Note that $\Delta\kappa_\infty$ is much smaller than the initial signal $\kappa(t)$ after the 248 nm photoexcitation (ca. 58 nS/cm for the trace shown in Fig. 3S(a)). If all anions present in the solution at $t = t_L$ were photoexcited, one would expect that at saturation, the conductivity signal induced by 1064 nm light would equal the initial signal from the electrons. The comparison of these signals suggests only 10% of the anions can be photoexcited by 1064 nm light. Assuming a typical electron scavenging rate constant of $(1-2) \times 10^{12}$ M$^{-1}$ cm$^{-1}$ (for $n$-hexane), [3,55] the concentration of the impurity which yields the interfering anion is < 0.5 μM. It was impractical to purify alkanes to < 500 ppb to exclude this impurity.

Fortunately, such a purification is not needed because at short delay times $t_L$ of the 1064 nm pulse, the weight of the signal from the anion is quite small (as very few electrons are scavenged at these short delay times) and the area $\Delta A$ under the fast component can be accurately determined (Fig. 2). To this end, the signal $\Delta\kappa(t)$ was fitted by a weighted sum of a Gaussian curve and its integral. The area under the Gaussian was taken as $\Delta A$ in eq. (4). The equilibrium conductivity signal $\kappa_0$ from the electrons at this delay time was estimated from $\kappa_0 = \kappa(t_L) - \kappa_{ion}$. The ratio of these two quantities yields the ratio $r$. Within the accuracy of our experiment, this ratio does change with the delay time $t_L$ of the 1064 nm laser pulse and remains linear with $J$ to at least $8 \times 10^{18}$ photon/cm$^2$. The $r/J$ slope (that according to eq. (5) and Appendix A in the Supplement equals $\langle\sigma\tau_t\rangle$) is $(2.5\pm0.1) \times 10^{-28}$ cm$^2$·s. Using the absorption cross section $\langle\sigma\rangle \approx 3.2 \times 10^{-17}$ cm$^2$ (for 1 μm light) of the electron in $n$-hexane at 295 K [56] as an estimate for the photodetachment cross section (Sec. IV.A), the "average" residence time $\bar{\tau}_t = \langle\sigma\tau_t\rangle/\langle\sigma\rangle$ of the electron of ca. 8.3 ps is obtained. For methylcyclohexane (in which $\langle\sigma\rangle \approx 3.3 \times 10^{-17}$ cm$^2$ for 1 μm light), [42] a similar measurement yields ca. 13 ps. We did not study other $n$-

14.

alkanes, cycloalkanes and their methyl derivatives because most of these hydrocarbons exhibit similar electron mobilities and activation energies to *n*-hexane and methyl-cyclohexane. Importantly we did not observe the 1064 nm photon induced signal (except for the signal from impurity anions) from highly branched alkane liquids that yield high-mobility electrons, such as *iso*-octane. Apparently, the detrapping time $\tau_t$ for the electrons in these liquids is too short and/or the photodetachment cross sections are too low to observe the photoinduced signal at 295 K, even at our large ( > 1 J/cm$^2$) laser fluence. According to Balakin et al.,[20] for *iso*-octane at 170 K, $\bar{\tau}_t$ is ca. 20 ps.

We turn now to the temperature dependencies of the conductivity signals for *n*-hexane (which has boiling and melting points at 342 K and 180 K, respectively). The data were obtained in a relatively narrow temperature range between 280 and 320 K (Figs. 3 and 4) since both the activation energy for electron migration [57,58] and the absorption spectrum [42] depend on the temperature, which complicates the analysis over a wide temperature range. To increase the conductivity signal in a smaller cell, the electrons were generated by photoionization of 5 μM anthracene. The anion of anthracene does not absorb in the near infrared,[59] and the same ratio $r$ was obtained with and without this photosensitizer, despite the severalfold increase in the conductivity signal $\kappa(t)$ and its shorter life time, due to the electron scavenging by the aromatic solute. The addition of anthracene actually *improves* the accuracy of the $\Delta A$ measurement as it decreases, via competitive electron scavenging, the yield of photoactive impurity anion interfering with the signal (see above). Figures 3 and 4 show the Arrhenius plots for the conductivity signal $\kappa_{ion}$ from the ions, the conductivity signal $\kappa_0$ from the electron (extrapolated to $t \rightarrow 0$ using an exponential fit), the area $\Delta A$ under the fast component of the photoinduced signal at the maximum fluence of 1064 nm photons, and the ratio $r$. All of these Arrhenius plots are linear within the experimental error.

The conductivity signals $\kappa_{ion}$ and $\kappa_0$ are given by the products of the ion/electron concentration and their mobility. Both of these quantities are temperature dependent [36,50] and the activation energies determined from the plots shown in Figure 3 are the sums of the activation energies for the photoionization yield $Y$ of the electrons/ions and the

15.

corresponding mobilities. [50] For the mobilities $\mu_\pm$ of molecular ions in *n*-hexane (see Fig. 4S for the literature data), the activation energies are 8.2±0.2 kJ/mol (for anions) and 11.3±0.1 kJ/mol (for cations). The activation energy for the sum $\mu_i$ of the ion mobilities is 9.4±0.2 kJ/mol. Given that the activation energy for $\kappa_{ion}$ is 18.2±0.8 kJ/mol, the activation energy for the ion (and, therefore, electron) yield $Y$ is ca. 8.8±1.0 kJ/mol. Similar estimates of 5.5±1 and 10±1 kJ/mol were obtained for the activation energy of bi-248 nm photon ionization of triphenylene in methylcyclohexane and cyclohexane, respectively. [50] Subtracting this activation energy from 32.6±0.4 kJ/mol obtained for $\kappa_e$, the activation energy for $\langle \mu_e \rangle$ is 23.8±1.4 kJ/mol (ca. 230 meV), which compares favorably with the time-of-flight estimate of 190 [37,39] to 230 meV. [4]

Figures 4(a) and 4(b) exhibit the temperature dependencies for the area $\Delta A$ and ratio $r$, respectively. This ratio, which is proportional to the detrapping time $\bar{\tau}_t = r/J$, obviously does not dependent on the photoionization yield (eq. (5)). The corresponding activation energy is 21±0.4 kJ/mol (ca. 200 meV). The activation energy for $\Delta A$ is ca. 2.7±0.5 kJ/mol, and the activation energy for the product $\mu_f \tau_f$ (which is proportional to $\Delta A/Y$, eq. (4)) is just -(6±2) kJ/mol. Another estimate for this energy (from the identity $\mu_f \tau_f \approx \langle \mu_e \rangle \bar{\tau}_t$) gives an estimate of -(4±2) kJ/mol. It appears that the activation energy is nearly zero and $\mu_f \tau_f \propto T$. Weak temperature dependencies for the mobility $\mu_f$ and the lifetime $\tau_f$ of quasifree electron is one of the tenets of the two-state model. [1] Our experiment validates these assumptions and supports the prevalent view that the steep temperature dependence for $\langle \mu_e \rangle$ originates through the phonon-assisted emission from traps. Note that in our estimates, it was assumed that the cross section for the electron detachment by 1064 nm photons does not change substantially over the narrow temperature range.

**IV. DISCUSSION**

We conclude that in room-temperature hydrocarbon liquids exhibiting low electron mobility $\langle \mu_e \rangle$ (ca. 0.1 cm$^2$/Vs), the mean residence time $\bar{\tau}_t$ of electrons in traps

16.

is shorter than 8-13 ps and the activation energy of detrapping (which should be close to the binding energy $E_t$ of the traps) is ca. 200 meV. Both of these estimates are close to those obtained by Mozumder [22,29] using the "quasiballistic" model for electron transport. As emphasized in Sec. II, the experimentally determined $\bar{\tau}_t$ actually provides an *upper* bound estimate for the detrapping time. As seen from our results, this upper bound is ca. 200 times smaller than the estimate obtained when $\tau_f \approx 310\pm100$ fs lifetime for the quasifree electron (estimated by Knoesel et al.) [36] is substituted into the two-state model. The lifetime-mobility product $\mu_f \tau_f$ for the quasifree electron (where the apparent $\mu_f$ might be reduced due to the incomplete velocity relaxation) [22,29] is ca. $10^{-12}$ cm$^2$/V and its mean path given by eq. (6) is $\Lambda_t \approx 40$ Å, which is well within the Onsager radius of 300 Å. Assuming that $\mu_f \approx 20$-50 cm$^2$/Vs (as suggested by electron Hall mobility [15,23] and $e^+$ Doppler effect mobility measurements), [26] we obtain $\tau_f \approx 20$-50 fs. Again, this is perfectly consistent with the estimate of ca. 30 fs obtained by Mozumder [29] using the "quasiballistic" model. On the other hand, our estimate is more than an order of magnitude lower than the damping time $\tau_d$ obtained from the THz spectrum of (quasifree?) electron in *n*-hexane. [35,36] The electron density in our study was as low or even *lower* than that in the THz study of Knoesel et al., [35,36] and the shortening of $\tau_f$ due to electron-electron scattering cannot be the reason for this discrepancy.

Before considering how to resolve this apparent contradiction, we re-examine the basic assumptions made in our analysis. In particular, the properties of trapped electrons in nonpolar liquids are not well understood. [1] For example, it has not yet been conclusively demonstrated that the electrons in shallow traps (which are involved in the conduction equilibria) are the same electrons which contribute to the absorption in the near and mid infrared. The arguments given below suggest that they are the same. *We argue that the entire absorption band of the electrons in liquid hydrocarbons originates from a bound-to-continuum transition from an electron "bubble" to the conduction band.*

## A. Absorption spectra and photodetachment cross sections.

Any further discussion requires a concrete model for the trapped electron in a nonpolar liquid. The simplest of such models is that of an electron trapped in a rectangular spherical potential well of depth $U$ and a hard core radius $a$. The electron would localize in such a well provided that $U > \pi^2 \hbar^2 / 8 m_e a^2$ [9] with a binding energy $E_t = E_t(U,a) < U$. This model, also known as the Wigner-Seitz model, [9,43,44,60] or the electron bubble model, [16,45-47] was originally suggested for electrons in liquid $^4$He. Since for this liquid $V_0 \approx 1$ eV (due to the negligible polarizability of He atoms) and $a \approx 17\text{-}20$ Å (due to the low surface tension), [16,47] the well is deep, supporting several bound-to-bound (bb) transitions (1s-1p [61,62] and 1s-2p) [45,61] in addition to bound-to-continuum (bc) transitions [63] from the ground 1s state. The bubble model accurately describes the pressure and temperature dependencies for the corresponding transition energies [45,47,61-63] and accounts for many other phenomena, such as the explosion of the electron bubbles in the acoustic field, [47,64] emission of the electrons from the bubbles across the surface, [17] sound wave generation, [65] vortex trapping, [66] etc.

This remarkable success prompted several workers [9,43,44,71] to use the bubble model for electrons in low-temperature, vitreous hydrocarbons which, like liquid $^4$He, also exhibit large positive $V_0$. According to these models, the entire absorption spectrum for such electrons originates through a bc transition from the ground 1s state. Metastable trapped electrons in glassy hydrocarbons can be observed using magnetic resonance techniques, such as EPR [67] and ESEEM. [68] According to these data, the 1s electron resides at the center of a spherical cavity with a radius $a$ of 3.4-3.6 Å. [67,68] This cavity is lined by CH$_3$ groups of the alkane molecules. The electron is weakly coupled (via magnetic dipole interaction) with ca. 20 methyl protons. [68] The preference for methyl protons is due to the higher polarizability of the C-H bonds in comparison to the C-C bonds. [69] To a first approximation, the depth $U$ of the potential well is given by $U \approx V_0 - E_{pol}$, where $E_{pol} = -(1 - \varepsilon^{-1}) e^2 / 2a$ is the Born polarization energy for a sphere of radius $a$. [9,43,60] Since $V_0$'s are large (0.6-1 eV), [70] the potential well is deep, and bound-to-bound (bb) transitions are possible. Consequently, there are two schools of thought

18.

concerning the electron spectra in the vitreous hydrocarbons: (i) that the entire spectrum is due to the bc transitions, [9,43,44,71] and (ii) that the bc transitions dominate only above a certain threshold energy $E_{bc}$ (ca. 1 eV); [10,72-74] below this energy, both the bc and bb transitions may occur. The latter situation occurs for solvated electrons in the polar media, where the potential wells are deep. [8,10]

By the well-known Wigner formula for a bc transition from the bound *s*-state to a free *p*-wave electron, [75] the cross section for $E \approx E_t$ is given by $\sigma_a(E) \propto (E - E_t)^{3/2}$. This result does not depend on the exact form of the potential provided that it decreases faster than $r^{-1}$ towards the bulk. McGrane and Lipsky [44] recently obtained the spectra of trapped electrons in glassy alkanes in the near and mid infrared and examined the low-energy "tail". Wigner's relation holds exactly, suggesting that the low-energy slope and, therefore, the *entire* spectrum of the trapped electron is due to the *bc* transitions. The analysis of these spectra using the spherical well model yields very similar estimates for the cavity radii *a* as those obtained using magnetic resonance spectroscopies. [67,68]

Following the photoexcitation, the electrons are promoted to the conduction band and for a brief time they become mobile, generating photocurrent and recombining with their parent cations. This recombination results in the bleaching of the trapped electron absorbance (and the decrease in its EPR signal) and also the luminescence. The argument has been given in the past [10] that the absorption spectrum cannot be entirely due to the bc transitions since the action spectra of photocurrent, photobleaching, and photoluminescence, at least for some vitreous hydrocarbons, are blue-shifted with respect to their absorption spectra. When the ratio of the corresponding cross sections is plotted against the photon energy $E$, it appears that the quantum yield $\phi_t$ of photodetrapping is a sigmoid function centered at $E = E_{bc}$ to the blue of the absorption maximum. [72,74,76] The constancy of $\phi_t$ for $E > E_{bc}$ suggests that $\phi_t \approx 1$. [72] The surprising aspect of these measurements is that the action spectra produced by different methods (absorption bleaching, [77,78] EPR, [79] conductivity, [72,74,76] and luminescence [77]) are not the same. [10] The similarity between the action spectra and absorption spectra was observed for some vitreous hydrocarbons (e.g., methylcyclohexane) [71] but not for others (e.g., 3-methylpentane). [72] The important point missing from the debates is that the detrapping *per se* does not generate any photocurrent since the conduction band electron has to escape the field of its parent cation first; that may require extra energy. Conversely, in order to recombine, this electron has to avoid being captured by its parent trap (whose relaxation is slow in the low-temperature solid), which also requires extra energy. Thus, the blue shift cannot be considered as a clear-cut evidence for the occurrence of a bb transition.

Given the relative success of the electron bubble model for solid hydrocarbons, it seems surprising that it has not been used to account for the properties of trapped electrons in *liquid* hydrocarbons. There are good reasons to believe that the latter species are also localized in voids. The experiments of Holroyd and co-workers [11,80] on the pressure dependence of $\langle \mu_e \rangle$ yield the volume change associated with electron trapping (see Sec. 6.2 of ref. 1). This quantity can be divided into a positive term corresponding to the cavity volume ($\propto a^3$) and a negative term ($\propto a^{-1}$) due to the electrostriction. This allows estimation of the cavity volume, which gives $a \approx 3.2\text{-}3.6$ Å, [1,11] close to the estimates obtained from the absorption spectra [44,60] and magnetic resonance data [67,68] in vitreous hydrocarbons. Molecular dynamics and path integral calculations for electrons in liquid ethane [81,82] and amorphous polyethylene [19,82] also predict that electrons reside in cavities of 5-7 Å in diameter.

The skepticism towards the validity of the bubble model for *liquid* hydrocarbons can be traced to Hammer et al., [83] where the stability of the electron bubbles was examined. The total energy $E_{tot}(R)$ of the bubble of radius $R$ relative to $V_0$ is given by $-E_t + \gamma R^2$, where the last term is the surface energy and $E_t = E_t\left(U \approx V_0 - E_{pol}, R\right)$. For the electron bubble to be stable, $E_{tot}(R) < 0$ should have a minimum at $R = a$. [46] This criterion is easy to satisfy in low temperature solids since $V_0$ is very positive, [60] however, it is not fulfilled in the room temperature liquid alkanes for which $V_0$ is close to zero (e.g., for *n*-hexane $V_0$ is ca. 100 meV at 295 K). [1,2,4,5] Hammer et al. [83] observed that the way to obtain the stable electron bubble is to assume that the local dielectric constant $\varepsilon$ is 50% higher than the average value of $\varepsilon \approx 2$ in the bulk. That the Born formula

20.

incorrectly estimates the polarization energy due to the neglect of the C-H bond polarization is a recurrent theme of many models of trapped electrons in vitreous alkanes (e.g., ref. 44). In the "microdipole" model of Kevan and co-workers, [69,84] the electron is stabilized via its interaction with the dipoles induced by the polarization of C-H bonds in methyl groups lining the solvation cavity. In this model, the self-consistency may only be achieved if one postulates for this C-H bond ca. 5 times greater polarizability than commonly assumed. It has been suggested [85] that the missing energy term might be the exchange interaction between the cavity electron and the valence electrons in the alkane molecules, which is neglected in the one-electron models. The need for such a term is clear from Fig. 5S(a), where the radial electron density for the ground state electron for $E_t = 180$ meV is plotted. The wavefunction extends well beyond the well radius to $r \approx 2a$. Such a diffuse electron density is peculiar only to solvated electrons in *shallow* traps. As there is considerable density in the region filled with the solvent molecules, a consistent microscopic theory should prescribe how the molecules interact with these molecules outside the void that makes the core of the electron bubble. Our density functional calculations [48] suggested that penetration of the electron density onto the methyl groups would indeed increase the polarizability of the C-H bonds in the manner suggested by Kevan et al. [69] consistently with the geometry suggested by their magnetic resonance data. [26,67] Although the bubble model obviously "works" in terms of predicting the optical spectra and the resulting parameters are consistent between various measurements, this model does not specify through which interactions the trapping potential $U$ originates (in other words, how the electron interacts with the bubble "wall") and what makes the cavity stable. Such are the limitations of this simple model.

What interests us most in relation to this study is whether the electron bubble model may account for the absorption spectra in liquid alkanes. Atherton et al. [42] obtained a series of such spectra for methylcyclohexane between 132 and 295 K (the melting point is 147 K). To our knowledge, these are the only data in the literature that are taken across a sufficiently wide range of wavelengths and temperatures. In Figure 5, the normalized spectra obtained by Atherton et al. [42] are fitted using the electron bubble model [9,44] assuming, following the previous discussion, that the bubble radii $a$ are not temperature dependent (we assumed $a \approx 3.4$ Å from the estimated cavity volume of 96±18 cm$^2$/mol [1] at 295 K) and varying the binding energy $E_t$ only. As seen from this figure, the model reproduces the spectral profiles quite well. The blue shift is due to the systematic increase in $E_t$ (and the well depth $U$) with the decreasing temperatures (Figs. 5(b) and 5(c)) The increase in $E_t$ is from 180 meV at 295 K to 345 meV at 132 K (the increase in $U$ is from 1.44 to 1.74 eV). The calculated molar absorptivity of the electron is 7x10$^3$ M$^{-1}$ cm$^{-1}$ (Fig. 5(a)) vs. the experimental estimate of 8.7x10$^3$ M$^{-1}$ cm$^{-1}$ for methylcyclohexane-d$_{14}$ at 295 K. [42]

The increase in the $U$ with the decreasing temperature is easy to rationalize, as $V_0$ rapidly increases in denser medium, whereas the polarization energy changes much less (this trend continues in solid methylcyclohexane, as suggested by the data of McGrane and Lipsky [44] for 77 K shown in Fig. 5(c)). The increase in $V_0$, as shown elsewhere, [86] can be rationalized using the model of Springett, Jortner, and Rice [46] in its formulation by Kevan et al. [87] This is illustrated in Fig. 5(c) using the experimental data for *n*-hexane [86] and calculated $V_0$ energies for methylcyclohexane. Following the decrease in $E_t$ with the increasing temperature, the activation energy for $\langle \mu_e \rangle$ should also decrease. The only data for the electron mobility across a temperature interval of comparable width to that explored by Atherton et al. [42] are for 3-methylpentane. [57] As the temperature decreases below 200 K, the electron mobility becomes lower than 10$^{-10}$ cm$^2$/Vs and the activation energy increases from 140 meV to 400 meV. [37] In very cold 3-methylpentane (100-160 K), the trapped electron can be observed using optically-detected EPR, [88] via its magnetic resonance line. [89] The width of this resonance line depends on the rate with which the electron samples (via hyperfine interaction) different orientations of C-H proton spins. The latter rate is determined by the rate of trap-to-trap hopping or repeated thermal emission and trapping. [88,89] From the linewidth analysis, it is possible to estimate the activation energy of these processes, which also turns out to be close to 400 meV. [88] Thus, the trends observed in Fig. 5(c) are paralleled by the trends observed in the electron mobility, linking the electron bubble and the two-state models together. As argued in Appendix B, this conclusion still holds if one assumes realistic variation in the binding energies of electron traps, as suggested by several authors. [28,31-33]

22.

In the bubble model, the quantum yield for electron photodetachment is unity. As mentioned above, there is some controversy concerning whether this yield is always unity across the entire absorption spectrum in glassy hydrocarbons. For electrons in liquid alkanes, bb transitions are unlikely given the smallness of the binding energies deduced from Figure 5: there are no other bound states apart from the ground $1s$ state. Furthermore, for methylcyclohexane, the quantum yield is constant ($\approx 1$) [71] across the entire absorption band of the electron even in a 77 K glass, and we are confident that this would be even more so in the liquid. Our estimates for $\bar{\tau}_t$ for $n$-hexane and methylcyclohexane are very similar, which suggests that this is also true for the former liquid. The calculations suggest that the absorption (photodetachment) cross section at 1060 nm should increase ca. 2 times as compared to 295 K (assuming the same trend for $E_t$ as in methylcyclohexane and $n$-hexane; see Fig. 5S(b)), i.e., from the data of Yakovlev and co-workers [6,41] it appears that at 180-185 K, $\bar{\tau}_t$ is ca. 420 ps for methylcyclohexane and ca. 200 ps for $n$-hexane (vs. ca. 13 ps and 8.3 ps, respectively, at 295 K). Our estimates for $E_t$ at 191 K and 295 K are 250 meV and 180 meV, respectively. Thus, the detrapping rate rapidly decreases with the increasing binding energy, in agreement with the two-state model.

We conclude that (i) the same electrons that contribute to the optical absorption spectrum are also involved in the equilibria which are responsible for electron conduction and (ii) the electron bubble model, despite its conceptual faults, provides a consistent description of trapped electrons in these solvents.

**B. Localization, trapping, and the two-state model.**

The considerations above suggest that our estimates for the mean residence time of the electrons in shallow traps should be reliable, at least within an order of magnitude, and various complications discussed in Sec. V.A do not change this conclusion. The absorption spectra of the electrons in liquid alkanes are consistent with a bc transition to the conduction band, and these spectra can be modeled quite well using the simple spherical well model with the parameters consistent between various independent measurements. The binding energies obtained from the spectral analysis are similar to the activation energies of electron detrapping. All of these quantities are temperature dependent. Our estimates appear to be in good agreement with those typically assumed in the two-state model (particularly, its recent formulation by Mozumder), [21,22,29,30] and may be regarded as further validation of this model.

On the other hand, as stated above, there appears to be no reasonable way to accommodate both our estimates for $\tau_t$ and the estimates of Knoesel et al. [36] for $\tau_f$ within the same two-state model postulating a thermal equilibrium between the localized and extended states. There are other problems with the estimate of $\tau_f \approx 310 \pm 100$ fs [36] (Sec. I) as it yields improbably high mobility and exceedingly long free path for the quasifree electron. Yet this estimate is not without a precedent:

Using femtosecond angle-resolved 2-photon photoemission (PE) spectroscopy, Ge et al. [90,91] recently studied the localization of $n=1$ surface-state electrons on thin layers of $n$-heptane covering Ag(111) at 120 K. The electrons are localized in the image potential well, in the direction normal to the surface; [91] still, such electrons may be delocalized in the transverse direction (in a similar fashion to the electrons on the surface of liquid $^4$He). [17,92] The parabolic band of the free electrons with $m_e^* \approx 1.2 m_e$ has been observed in the PE spectra. The lifetimes for these delocalized electrons range from 800 fs to 200 fs, depending on the in-plane component $k_\parallel$ of the wave vector (that changed from 0 to 0.23 Å$^{-1}$, respectively). [90] This time scale is similar to the estimate of the electron scattering time by Knoesel et al. [36] On the other hand, the energetics of electron localization on the surface is much different from that in the bulk liquid: e.g., according to Ge et al., [90,91] the localized state (a small polaron) has an energy just 10 meV lower than the bottom of the conduction band.

A relatively slow rate of electron trapping would also be consistent with theoretical modeling. The theorists have long striven to simulate the electrons in nonpolar liquids (mainly, helium [93-100] and rare gases, [97,100-103] and small hydrocarbons, such as methane and ethane) [81,82] and disordered dielectric solids (such as amorphous polyethylene) [18,19,82,104,105] using density functional theories, [94,95,102] path integral Monte-Carlo methods, [81,96,97,101] RISM-polaron theories, [94,101-103] mixed quantum chemical -

24.

molecular dynamics (QM/MD) calculations, [18,19,81,96,99,100] and, most recently, *ab initio* [104,105] and Car-Parrinello calculations. [105] These theories suggest a cavity electron as the lowest state for electron in helium, [106] ethane [81,82] and polyethylene [19,82] and delocalized, quasifree electron in Ar, Xe, and methane. [81,82,106] However, with the exception of QM/MD, [19,82,96,97] these methods are unsuitable to study the dynamics of electron localization and/or detrapping. Only recently did the theory start to address such issues. [82,93] Several important clues were obtained:

For electrons in near critical (309 K), dense (reduced density of 0.9) liquid-like helium, Space and Coker [93] observed that the localization/trapping are bimodal. The fast process takes just 50-100 fs, over which the electron is localized in a nodeless state; this rapid localization is followed by a slower relaxation that takes 200-700 fs, in which a solvation cavity (the electron bubble) gradually emerged. During this relaxation, the electron rapidly hopped between the adjacent density fluctuations (the proto-cavities), in contradiction to the premises of the two-state model. The relatively long time scale for electron trapping in liquid $^4$He is supported by experiments of Silver and co-workers [107] yielded an estimate of 300 fs for the electron trapping time at 1.4 K (there are authors suggesting even longer times). [108] The model of Rosenblit and Jortner [65] gives ca. 8-10 ps expansion times for the spherical cavity during its thermalization. Thus, the time scale for electron trapping appears to be nearly as long or even longer than the time for the solvation of electrons in water (250-300 fs) [109] and ammonia (200±50 fs). [110] For amorphous polyethylene, the calculations of Cubero and Quirke [19] suggest that the trapping (as opposed to the initial localization) takes picoseconds. The same calculation indicates that detrapping is relatively fast (tens of picoseconds) even for relatively deep traps ($\langle E_t \rangle \approx 350$ meV), [19,82] as the detrapping is driven by a large increase in the entropy. [19] These estimates are similar to ours for *n*-hexane and methylcyclohexane at 295 K (Sec. IV) as well as those of Balakin and Yakovlev for these two hydrocarbons at 180-185 K. [41]

The QM/MD calculations indicate that the two-state model is fundamentally flawed, as it makes no distinction between electron localization and the subsequent trapping. The latter involves the reorganization of solvent molecules around the excess

25.

electron. Such a process takes considerably longer time than the initial localization (that occurs in a few tens of femtoseconds, that is, a few C-H vibration cycles).[22,29] While modifications of the two-state model to implement such a distinction between the localization and relaxation/trapping are possible (see, for example, Sakai et al.)[111] many simplifying assumptions have to be made, without much justification. The two-state models of electron transport are successful only so far as the exact description of the localization/trapping process is not necessary. This is typically the case in the conductivity studies since the variation in $\langle \mu_e \rangle$ as a function of the solvent structure, temperature, etc. is mainly determined by the rate of thermal emission from traps rather than the intricate details of localization and trapping. Another reason for the success of these models is that the intermediate states of the trapping process seem to contribute little to the overall electron conduction.

### C. Reinterpretation of terahertz spectra.

While the two-state model may be deficient in more than one way, we believe that the apparent conflict of this model and the properties of quasifree electron obtained from the THz experiments of Knoesel et al.[36] has another explanation. Basically, no proof has been provided that the observed THz signal is indeed from the quasifree electron. Rather, the experiments suggest that (i) the THz radiation is absorbed by a Drude oscillator and (ii) the concentration of these Drude oscillators correlates with the overall yield of the excess electrons in a hydrocarbon. Thus, any electron species that behaves like a Drude oscillator with a damping time $\tau_d \approx 310 \pm 100$ fs would fit the observations. A circular argument was then given by Knoesel et al.:[35,36] a Drude oscillator with an effective mass $\approx m_e$ would exhibit a drift mobility > 10 cm$^2$/Vs; this large mobility points to the quasifree electron; this identification justifies the estimate for the oscillator's mass. The problem with this argument is that the electron bubble may also be considered as a massive Drude oscillator. There is a well-known precedent for such a treatment: The absorption of microwave (GHz) radiation by damped oscillations of electron bubbles trapped in the clamping electric field under the surface of liquid $^4$He has been observed as early as 1972 (see Sec. IV.A of ref. 17). The widths of the resonance lines perfectly agree with the predictions of the Drude model for a massive oscillator. For an electron bubble,

26.

the reduced mass $M$ is 1/2 of the mass of displaced liquid; [17,28] for *n*-hexane this would be ca. 30-100 a.m.u. The mobility $\mu_t$ of such an oscillator is given by $\mu_t = e\tau_d/M$. Substituting $\tau_d \approx 300$ fs in this formula, we obtain ca. $(3\text{-}10) \times 10^{-3}$ cm$^2$/Vs, which is a reasonable estimate for trapped-electron mobility. [28] Similar estimates were obtained using a semihydrodynamic theory for the electron bubble migration (see eqs. (4) and (5) in ref. 79). This theory suggests that the damping time $\tau_d \approx 3M^{1/2}/\left(8\langle n\rangle a^2 (2\pi kT)^{1/2}\right)$ would be independent of the viscosity of the liquid (in this equation $\langle n \rangle$ is the average number density). Since at any time most of the excess electrons in alkane liquids are trapped as electron bubbles, the THz signal from such *trapped* electrons can easily swamp the signal from the quasifree electrons, whose equilibrium fraction in *n*-hexane is only $3 \times 10^{-3}$ at 295 K. It looks probable that the THz signal observed by Knoesel et al. [35,36] is from the electron bubbles rather than a free carrier plasma.

Furthermore, as seen from Fig. 7 in ref. 36, the Drude oscillator model provided a rather poor fit at the lower end of the observation range, for frequencies $\omega/2\pi$ of 0.4-0.6 THz. Such deviations may originate from the effect of bubble vibrations on its oscillation in the electric field since the resonance frequencies of these vibrations fall into this sub-THz range. Gross and Tung-Li [112] and Celli et al. [113] gave detailed analyses of the vibrational modes for the electron bubbles in liquid helium; their theories can be readily adapted to other liquids. Using the expression for the frequency $\omega_0$ of the breathing ($l = 0$) mode of the bubble obtained by Gross and Tung-Li, [112] $\omega_0^2 = 4\gamma/M$ (similar to the Rayleigh formula for gas bubbles) and assuming the bulk value of $\gamma = 2 \times 10^{-2}$ J/m$^2$ for the surface tension in *n*-hexane, [83] one obtains an estimate of $\omega_0/2\pi \approx 0.1\text{-}0.2$ THz. The frequencies of $l = 2$ and $l = 3$ modes are ca. 1.4 and 7 times higher, respectively. [112] If, as suggested here, Knoesel et al. [35,36] have observed the effect of bubble vibrations on the THz conductivity, their experiment would be the first demonstration of such vibrations in any liquid, fulfilling the theoretical predictions made almost four decades ago!

Interestingly, Knoesel et al. [114] did not observe a significant increase in the electron signal when (trapped) electrons were photoexcited by a short 800 nm laser pulse and photogenerated "quasifree electrons" probed by a coincident THz pulse, though

27.

under their excitation conditions a 2-fold increase in the THz signal might have been expected. This preliminary result argues against the long trapping times for the quasifree electrons. Since the cycle of the THz wave was ca. 2 ps ($>> \tau_d$), it is possible that the free electrons decayed *before* detection. However, the simplest way to interpret this result is to assume, following the suggestion made above, that the localization of quasifree electrons was rapid (20-50 fs) and the THz signal originated from the oscillations of (thermalized) electron bubbles instead.

**V. CONCLUSION.**

Photoexcitation of trapped electrons in liquid *n*-hexane and methylcyclohexane at 280-320 K has been studied (Sec. IV). The absorption of 1064 nm light promotes the electron from the trap (the electron bubble) back into the conduction band of the liquid, increasing the conductivity by many orders of magnitude before the photogenerated free electron is localized and trapped. From the analysis of the data (loosely based on the approach of Yakovlev and co-workers [6,20,41] further developed in this study), the mean residence time $\tau_t$ of the electrons in traps has been estimated to be ca. 8.3 ps for *n*-hexane and ca. 13 ps for methylcyclohexane (at 295 K). The rate of detrapping decreases with decreasing temperature with an activation energy of ca. 200 meV (280-320 K), whereas the lifetime-mobility product $\mu_f \tau_f$ for quasifree electrons scales linearly with temperature in the specified range. The estimated localization time of the quasifree electron is 20-50 fs and its mean path is ca. 40 Å. Our estimates for the electron localization and detrapping times are in agreement with the "quasiballistic" model of Mozumder. [21,22,29,30]

On the other hand, this localization time is significantly shorter than 310±100 fs obtained for the scattering time of the free electron plasma in room-temperature *n*-hexane and cyclohexane using time-domain THz spectroscopy. [35,36] This conflict goes to the very foundations of the electron transport theories for nonpolar liquids, as it cannot be resolved within the standard (thermal equilibrium) two-state model. We suggest, however, that there is, actually, no such conflict: The THz signal originates from the oscillations of the electron bubbles in the THz electric field rather than from the dynamics of the free-electron plasma. Vibrations of these bubbles may be responsible for the deviations from the Drude behavior observed for frequencies below 0.4 THz (Sec. V.C). Other known properties of trapped electrons in liquid hydrocarbons can also be consistently accounted for using the cavity model (Sec. V.A).

While both this cavity model and, more generally, the equilibrium two-state model are obviously incomplete (Sec. V.B), it is remarkable how well these two models can rationalize various properties of electrons in hydrocarbon liquids, in a consistent way. This suggests that more advanced microscopic models of electron transport that are presently being developed should retain several crucial features of these models. Specifically, the trapped electron, regardless of the exact manner in which it interacts with the molecules lining the solvation cavity (electron exchange, polarization of C-H bonds, etc.), should behave more-or-less like a particle in a box (that is, a deformable, aspherical, vibrating cavity) of ca. 7 Å in diameter, with a binding energy of 180-200 meV (at 295 K). This energy should increase with the liquid density and decrease with the temperature, following the same trend as $V_0$. The electron localization and trapping, regardless of how these two processes exactly proceed, should be such that the free electron loses the momentum in a few tens of femtoseconds and the trapping is fully over in a picosecond. There should be a dynamic equilibrium between the trapped and the free electrons; the characteristic time for the thermal emission of the trapped electron should be tens of picoseconds. Any theory that yields these patterns would automatically provide the features that make the electron bubble and two-state models so remarkably successful.

## VI. ACKNOWLEDGEMENT.

IAS thanks Drs. Yu. A. Berlin and R. A. Holroyd and Profs. S. Lipsky, E. Knoesel, P. F. Barbara, A. Mozumder, L. D. A. Siebbeles, and K. Itoh for useful discussions. This work was supported by the Office of Science, Division of Chemical Sciences, US-DOE under contract number W-31-109-ENG-38.

**Figure captions.**

**Fig. 1.**

Time-resolved d.c. photoconductivity signals from room-temperature $N_2$-saturated neat *n*-hexane ionized by two 248 nm photons at $t = 0$. The solid line (coordinate scale to the left) is the conductivity signal $\kappa - \kappa_{ion}$ from the electrons. The signals $\Delta\kappa$ observed after the action of a 6 ns fwhm, 1.5 J/cm$^2$ pulse of 1064 nm photons are plotted to the right. Different traces correspond to different delay times of the 1064 nm pulse, from 50 ns to 600 ns. The empty circles are the weights of the fast component juxtaposed onto the $\kappa - \kappa_{ion}$ trace. See Sec. III for more explanation. The color version of this plot and the least squares fits for the $\Delta\kappa$ kinetics are given in Fig. 2S(a) in the Supplement. The data for methylcyclohexane are shown in Fig. 2S(b).

**Fig. 2.**

(a) Same as in Fig. 1, under slightly different excitation conditions. At short delay time $t_L \approx 45$ ns, the signal $\Delta\kappa$ (filled circles; to the right) is almost entirely from the 1064 nm photoexcitation of trapped electrons. The $\kappa - \kappa_{ion}$ signal (to the left and to the top), which is proportional to the total concentration of electrons in the photoionized solution, decays exponentially (the light curve is the least squares fit) at this low electron concentration (ca. 10 nM), in a scavenging reaction with an impurity. (b) The 1064 nm photon fluence dependence for the ratio $r$ given by eq. (5) determined at $t_L \approx 45$ ns (as seen from Fig. 1, this ratio does not change with the delay time of the 1064 nm pulse). The plot is linear, and the slope gives the mean product $\langle \sigma \tau_t \rangle$ of the photodetachment cross section $\sigma$ times the residence time $\tau_t$ for electrons in traps.

**Fig. 3.**

Arrhenius plots for the conductivity signals from (a) electrons (obtained from exponential extrapolation of $\kappa(t) - \kappa_{ion}$ to $t = 0$) and (b) ions ($t \approx$ 1-3 μm signal) in biphotonic excitation of 5 μm anthracene in *n*-hexane (all experimental conditions except for the temperature were the same for all runs). Several series of the data are plotted together to illustrate the scatter.

**Fig. 4.**

Arrhenius plots for (a) $\Delta A$, the area under the 1064 nm photon induced conductivity signal $\Delta\kappa$ (for $t_L \approx 50$ ns and a photon fluence of 1.5 J/cm$^2$; otherwise, the same excitation conditions as in Fig. 3) and (b) the ratio $r$ (eq. (5)). Several series of data obtained at 280 to 320 K are plotted together to illustrate the scatter.

**Fig. 5.**

(a) *Empty circles:* the absorption spectrum of solvated/trapped electron in pulse radiolyzed neat methylcyclohexane-$d_{14}$ at 295 K (after Atherton et al.);[42] $\varepsilon$ is the molar absorptivity (ca. 8.7x10$^3$ M$^{-1}$ cm$^{-1}$ at 1 μm). The lines are theoretical spectra calculated using the spherical well model for *(solid line)* a fixed radius $a$=3.36 Å (volume of 96 cm$^3$/mol)[1] and $E_t$=180 meV ($U$=1.44 eV) and *(dashed line)* $a$=2.6 Å and $E_t$=170 meV ($U$=2.1 eV). (b) *Symbols:* experimental spectra of the electron in neat methylcyclohexane-$h_{14}$;[42] the temperatures in K are indicated in the legend. The lines are normalized theoretical curves obtained using the spherical well model for temperature-independent $a$=3.36 Å (non-normalized traces are given in Fig. 5S(b)). The optimum parameters $E_t$ (filled circles; to the left) and $U$ (filled squares; to the right) are shown as a function of reciprocal temperature in part (c) of the figure. The empty circle shows $E_t$ for methylcyclohexane glass at 77 K.[44] The solid line drawn through the points is a guide for the eye. Also shown are the $V_0$ energies for *n*-hexane (empty triangles; to the left) and theoretical estimates of this energy using the approach of ref. 87. Molecular polarizability of 13.1 Å$^3$ and Wigner-Seitz radius of 2.235 Å for methylcyclohexane were assumed in the calculation. The increase in $E_t$ and $U$ with decreasing temperature follows the concomitant increase in the $V_0$.

**References.**

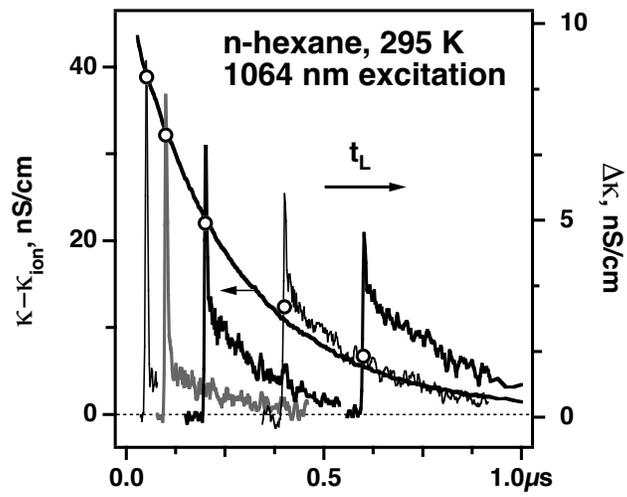

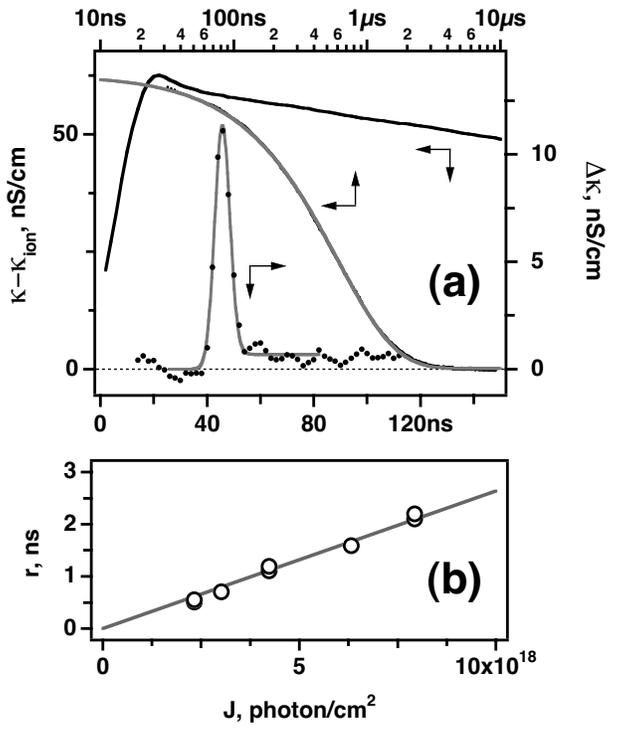

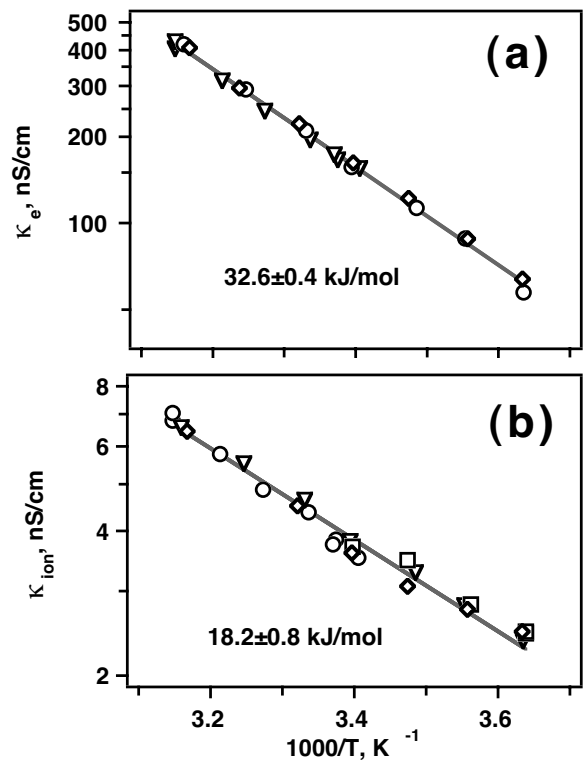

Figure 4; Shkrob & Sauer

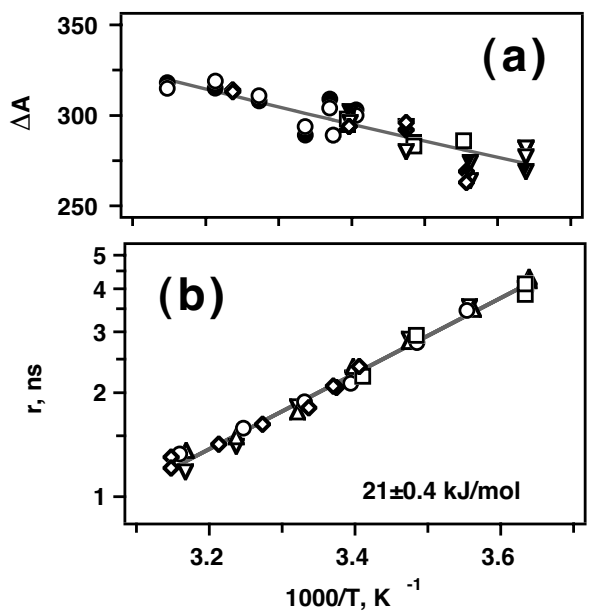

Figure 5; Shkrob & Sauer

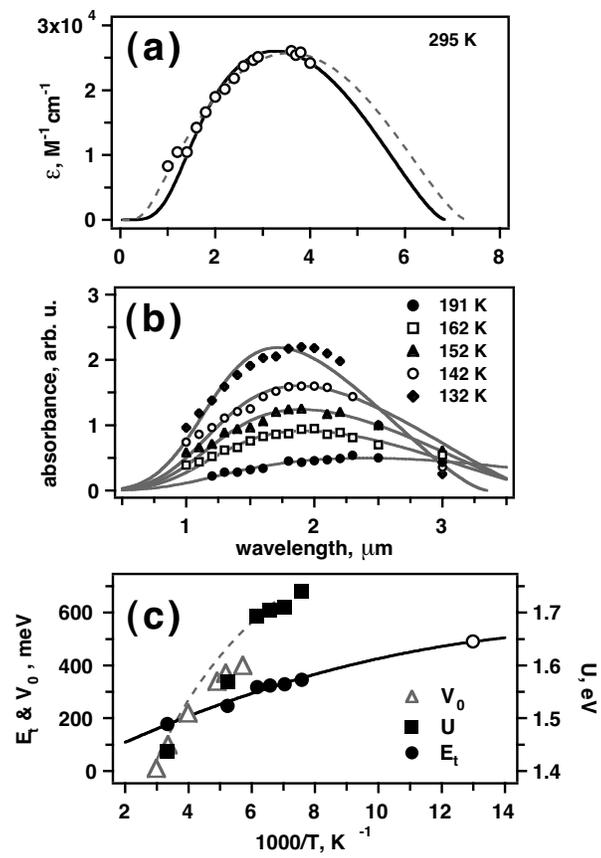

**Photo-Stimulated Electron Detrapping and the Two-State Model for Electron Transport in Nonpolar Liquids.**

*Ilya A. Shkrob and Myran C. Sauer, Jr.*

Chemistry Division , Argonne National Laboratory, 9700 S. Cass Ave., Argonne IL 60439

*Tel* 630-2529516, *FAX* 630-2524993, *e-mail:* shkrob@anl.gov.

# Supporting Information.

**Appendix A. Derivation of basic formulas.**

We first consider interconversion between two species, electron-1 and electron-2, with mobilities $\mu_1$ and $\mu_2$, that exist in the state of thermodynamic equilibrium. Let $e_1^0$ and $e_2^0$ be their equilibrium concentrations and $k_{12}$ and $k_{21}$ be the rate constants for $1 \to 2$ and $2 \to 1$ reactions, respectively. The equilibrium constant is given by $K = k_{12}/k_{21}$ and the equilibrium conductivity $\kappa_0$ by

$$\kappa_0 = F\left(\mu_1 e_1^0 + \mu_2 e_2^0\right) = F\,\frac{\mu_1 + K\mu_2}{1 + K}\, e^0, \tag{A1}$$

where $e^0 = e_1^0 + e_2^0$ and $F$ is the Faraday constant. Consider a laser pulse with time-dependent irradiance $J(t)$ that photoexcites electron-2 and converts it to electron-1. Let $\sigma_2$ be the cross section of this photoconversion. Since $e^0 = e_1 + e_2$ does not change during the photoexcitation, $de_2/dt = -de_1/dt$. The concentration of electron-1 is given by a kinetic equation

$$de_1/dt = -k_{12}\, e_1 + \left[k_{21} + \sigma_2 J(t)\right]\, e_2. \tag{A2}$$

In eq. (A2), it has been assumed that the photoexcitation of electron-2 yields the final state ("hot" electron) that very rapidly converts to electron-1. We will also assume that the lifetime-mobility product of this state is much smaller than this product for electron-1. We first examine the case when the irradiance is sufficiently small and the change $\Delta e_1(t) = e_1(t) - e_1^0 \ll e_2^0$. Then, we may approximate

$$d\Delta e_1/dt = -k_{12}\left(e_1^0 + \Delta e_1\right) + \left[k_{21} + \sigma_2 J(t)\right]\left(e_2^0 - \Delta e_1\right)$$
$$\approx -(k_{12} + k_{21})\Delta e_1 + \sigma_2 J(t) e_2^0 \,, \tag{A3}$$

where we used the fact that $de_1^0/dt = -k_{12}e_1^0 + k_{21}e_2^0 = 0$ at the equilibrium. Since before and after the laser pulse $\Delta e_1 = 0$, the integration of both sides of eq. (A3) from $t = -\infty$ to $t = +\infty$ yields the identity

$$\int_{-\infty}^{+\infty} dt \; \Delta e_1 \approx \sigma_2 e_2^0 J / (k_{12} + k_{21}), \tag{A4}$$

where $J = \int_{-\infty}^{+\infty} dt \; J(t)$ is the total photon fluence. During the photoexcitation, the conductivity signal from the electrons $\kappa(t)$ is given by

$$\kappa(t) = F\left(\mu_1 e_1 + \mu_2 e_2\right) = \kappa_0 + F(\mu_1 - \mu_2)\Delta e_1,$$

and the ratio $r = \Delta A/\kappa_0$, where $\Delta A = \int_{-\infty}^{+\infty} dt \left[\kappa(t) - \kappa_0\right]$ is the area under the photoinduced conductivity signal. This ratio is, therefore, given by

$$r = \kappa_0^{-1}(\mu_1 - \mu_2)\int_{-\infty}^{+\infty} dt \; \Delta e_1 \; . \tag{A5}$$

Combining eqs. (A4) and (A5), one obtains

$$r \approx \frac{(\mu_1 - \mu_2)\tau_1 \tau_2}{(\mu_1 \tau_1 + \mu_2 \tau_2)(\tau_1 + \tau_2)} \; \sigma_2 \tau_2 J, \tag{A6}$$

where $\tau_1 = k_{12}^{-1}$ and $\tau_2 = k_{21}^{-1}$ are the lifetimes of electron-1 and electron-2, respectively. For $\mu_1 >> \mu_2$ and $\tau_2 >> \tau_1$, the first factor on the right hand side of eq. (A6) is unity, and $r \approx \sigma_2 \tau_2 J$. It is easy to see that for $\mu_1 >> \mu_2$ and $\tau_1 << \tau_2, [\sigma_2 J(t)]^{-1}$ the latter equation holds in general. Indeed, in such a case, a quasi-stationary condition for electron-1 may be assumed, and $e_1$ can be determined from eq. (A2), by letting $de_1/dt = 0$ and equating $e_2^0 \approx e^0$, so that $e_1 \approx (\sigma_2 J(t) + k_{21})e^0/k_{12}$ and

$$\Delta\kappa(t) = \kappa(t) - \kappa_0 \approx (\mu_1 - \mu_2)\sigma_2 \tau_1 J(t). \tag{A7}$$

Integrating the latter equation gives

$$r \approx \frac{(\mu_1 - \mu_2)\tau_1}{\mu_1\tau_1 + \mu_2\tau_2} \sigma_2\tau_2 J, \tag{A8}$$

which is equivalent to eq. (A6) for $\tau_2 \gg \tau_1$. For $\mu_1 \gg \mu_2$, we obtain $r \approx \sigma_2\tau_2 J$. Identifying "electron-1" with the quasifree electron and "electron-2" with a trapped electron, eq. (5) is obtained. Thus, eq. (5) is correct under very general assumptions, provided that $\sigma_2 J(t)\tau_1 \ll 1$ during the photoexcitation.

A similar approach can be used to obtain the general result for multiple electron traps. To this end, we introduce $\sigma_m$, the photodetachment cross section, and $\tau_m$, the residence time, of the electron residing in a trap of kind *m*. Below we demonstrate that $r \approx \langle \sigma_m\tau_m \rangle J$, where the averaging $\langle ... \rangle$ is taken over the equilibrium concentrations of all trapped electrons. While this formula is intuitively obvious, the derivation is lengthy and cumbersome. Furthermore, it is easy to see that this formula is correct only for *small* laser fluences. When the fluence is large, the equilibria between the electrons in different traps can be shifted during the pulse and a phenomenon similar to saturation manifests itself. This phenomenon has been observed experimentally, [40] when polar molecules were added to *n*-hexane, to create new kinds of electron traps in the solution.

Let $e_f$ be the concentration of quasifree electrons which are in the state of equilibrium with several electron traps; the concentration of electrons in these traps is denoted by $e_m$:

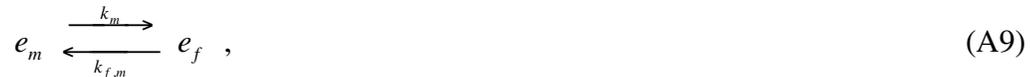

$$e_m \underset{k_{f,m}}{\overset{k_m}{\rightleftarrows}} e_f, \tag{A9}$$

In reaction (A9), $k_m = \tau_m^{-1}$ is the reciprocal residence time $\tau_m$ of the electron in a given trap and $k_{f,m}$ is the rate of descent of the quasifree electron into this trap. The kinetic equations, during the photoexcitation of trapped electrons, are given by

$$\frac{de_m}{dt} = k_{f,m}e_f - [\sigma_m J(t) + k_m]e_m, \tag{A10}$$

$$\frac{de_f}{dt} = \sum_n [\sigma_n J(t) + k_n]e_n - k_f e_f \tag{A11}$$

where $k_f = \tau_f^{-1} = \sum_n k_{f,n}$ is the total rate constant of electron trapping and $\tau_f$ is the lifetime of the quasifree electron. For $e_f \ll e_m$ and $\sigma_m J(t) + k_m \ll k_f$, we can assume stationary conditions for the quasifree electron during the photoexcitation, obtaining

$$e_f(t) = \sum_m e_f^{(m)}(t), \qquad (A12)$$

where

$$e_f^{(m)}(t) = \frac{\sigma_m J(t) + k_m}{k_f} e_m(t). \qquad (A13)$$

At equilibrium, the concentrations $e_m^0$ of the electrons in the corresponding traps are given by

$$e_m^0 = \frac{k_{f,m} e_f^0}{k_m} = k_f P_m \tau_m e_f^0 \qquad (A14)$$

where $e_f^0$ is the equilibrium concentration of quasifree electrons and $P_m = k_{f,m}/k_f$ is the partition coefficient. One can formally consider an equilibrium between the electrons in different traps, reaction

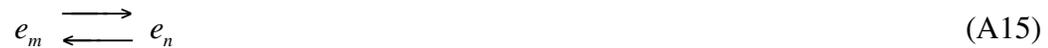

$$e_m \rightleftharpoons e_n \qquad (A15)$$

with the equilibrium constant

$$K_{m \to n} = \frac{e_n^0}{e_m^0} = \frac{P_n \tau_n}{P_m \tau_m} \qquad (A16)$$

and the equilibrium fraction $f_m^0$ of trapped electrons given by

$$f_m^0 = \frac{e_m^0}{\sum_n e_n^0} = \frac{P_m \tau_m}{\sum_n P_n \tau_n}. \qquad (A17)$$

Prior to the laser pulse, eqs. (A12) and (A13) simplify to

$$e_f^0 = \sum_m e_f^{(m),0}, \qquad (A18)$$

and

$$e_f^{(m),0} = \frac{k_m e_m^0}{k_f}, \tag{A19}$$

where the subscript "0" corresponds to the equilibrium concentrations. Assuming that the mobility $\mu_f$ of the quasifree electron is much greater than the mobility of electrons in traps, $\mu_f \gg \mu_m$, we have $\kappa_0 = F\mu_f e_f^0$. For the ratio $r$, we obtain

$$r = \int dt \ [\kappa(t) - \kappa_0]/\kappa_0 = \int dt \ [e_f(t) - e_f^0]/e_f^0, \tag{A20}$$

where the integration is from $t = -\infty$ to $t = +\infty$. Substituting eqs. (A12), (A13), (A18), and (A19) into eq. (A20) gives

$$r = \sum_m \int dt \ \varepsilon_m(t), \tag{A21}$$

where

$$\varepsilon_m(t) = \frac{e_f^{(m)}(t) - e_f^{(m),0}}{e_f^0} = P_m \left\{ [1 + \sigma_m \tau_m J(t)] \frac{e_m(t)}{e_m^0} - 1 \right\} \tag{A22}$$

We proceed to estimate the integral $\int dt \ \varepsilon_m(t)$. Substituting eq. (A12) into eq. (A10), a system of linear equations for $\{e_m\}$ is obtained,

$$\frac{de_m}{dt} = -[\sigma_m J(t) + k_m]e_m + P_m \sum_n [\sigma_n J(t) + k_n]e_n. \tag{A23}$$

As previously, we assume that $\Delta e_m(t) = e_m(t) - e_m^0 \ll e_m^0$ and expand the integral

$$\int dt \ \varepsilon_m(t) \approx P_m \left\{ \sigma_m \tau_m J + (e_m^0)^{-1} \int dt \ \Delta e_m(t) \right\}, \tag{A24}$$

simultaneously recasting eq. (A23) as

$$\frac{d\Delta e_m}{dt} \approx -k_m \Delta e_m + P_m \sum_n k_n \Delta e_n - J(t) \left[ \sigma_m e_m^0 - P_m \sum_n \sigma_n e_n^0 \right]. \tag{A25}$$

Since $\int dt \ (d\Delta e_m/dt) = 0$, the integration of both sides of eq. (A25) yields

$$\int dt \ \Delta e_m = P_m \tau_m \Omega - \left[ \sigma_m \tau_m e_m^0 - P_m \tau_m \sum_n \sigma_n e_n^0 \right] J, \tag{A26}$$

where

$$\Omega = \sum_n k_n \int dt \ \Delta e_n . \quad (A27)$$

Since $\sum_m \Delta e_m = 0$, $\sum_m \int dt \ \Delta e_m = 0$. Summing up all eqs. (A26) and equating the resulting formula to zero, we obtain

$$\frac{\Omega}{J} = \frac{\sum_n \sigma_n \tau_n e_n^0}{\sum_n P_n \tau_n} - \sum_n \sigma_n e_n^0 . \quad (A28)$$

Substituting eq. (A28) into eq. (A16) and using eq. (A17), we obtain

$$(Je_m^0)^{-1} \int dt \ \Delta e_m = -\sigma_m \tau_m + \sum_n (\sigma_n \tau_n) f_n^0 , \quad (A29)$$

from which, by way of eqs. (A21) and (A24), the final result is obtained

$$r \approx \sum_m \sigma_m \tau_m f_m^0 \ J \equiv \langle \sigma_m \tau_m \rangle \ J \quad (A30.a)$$

proving the assertion given above. Observe that partition coefficients $P_m$ do not enter expression (A30.a) explicitly; the apparent electron mobility $\langle \mu_e \rangle \approx \mu_f \tau_f \langle \tau_m^{-1} \rangle$ also does not depend on these partition coefficients explicitly.

Eq. (A30) was obtained for a system in which the electrons in different traps equilibrate via the thermal emission and subsequent trapping. The formula can be further generalized if other ways exist for electrons to equilibrate between the traps, by changing the definition of the residence time for the electron in a given trap. Consider equilibrium reactions (A15), with forward and backward rate constants $k_{n \to m}$ and $k_{m \to n}$ for transfer of electrons between the corresponding traps (in addition to reactions (A9)) that does not involve thermally assisted electron detachment. It is easy to demonstrate that in such a case, the ratio $r$ is given by expression

$$r \approx \langle \sigma_m \tau_m' \rangle J \quad (A30.b)$$

in which the residence times $\tau_m'$ are the solutions of a system of linear equations

$$\sum_n Q_{mn} \tau_n' = 1, \quad (A31)$$

with the matrix elements given by

$$Q_{mn} = \left(k_m + \sum_{l \neq m} k_{m \to l}\right)\delta_{mn} - k_{n \to m}. \tag{A33}$$

The derivation of eq. (A30.b) is straightforward. Since eq. (A11) still holds in this more general case, $k_f e_f^0 = \sum_m k_m e_m^0$ and $k_f e_f = \sum_m [k_m + \sigma_m J(t)] e_m$, and eq. (A23) can be rewritten as

$$\frac{de_m}{dt} = P_m \sum_m [k_m + \sigma_m J(t)] e_m - \sigma_m J(t) e_m - \sum_n Q_{mn} e_n. \tag{A34}$$

At equilibrium, the right hand side of eq. (A34) equals zero and

$$P_m \sum_n k_n e_n^0 = \sum_n Q_{mn} e_n^0. \tag{A35}$$

Eq. (A21) is still correct and, retaining only linear terms, we can write

$$\varepsilon_m(t) = \frac{[k_m + \sigma_m J(t)] e_m - k_m e_m^0}{k_f e_f^0} \approx \frac{\sigma_m e_m^0 J(t) + k_m \Delta e_m}{\sum_n k_n e_n^0}, \tag{A36}$$

so that

$$r \approx \frac{\sum_n \sigma_n e_n^0 J + k_n \int dt \; \Delta e_n}{\sum_n k_n e_n^0}. \tag{A37}$$

Retracing the steps made to derive eqs. (A25) and (A26), the identity

$$P_m \left\{\sum_n \sigma_n e_n^0 J + k_n \int dt \; \Delta e_n\right\} = \sigma_m e_m^0 J + \sum_n Q_{mn} \int dt \; \Delta e_n. \tag{A38}$$

is obtained. Dividing this equation by eq. (A35) and using eq. (A37), we obtain that for any $m$

$$r = \frac{\sigma_m e_m^0 J + \sum_n Q_{mn} \int dt \; \Delta e_n}{\sum_n Q_{mn} e_n^0}, \tag{A39}$$

or, in the matrix form,

$$r \ \mathbf{Q} \ \mathbf{e}^0 \ = \ \sigma \bullet \mathbf{e}^0 \ J + \ \mathbf{Q} \ \int dt \ \Delta \mathbf{e}(t) \ , \tag{A40}$$

so that

$$r \ \mathbf{e}^0 \ = \ \mathbf{Q}^{-1} \ \sigma \bullet \mathbf{e}^0 \ J \ + \ \int dt \ \Delta \mathbf{e}(t). \tag{A41}$$

Adding the rows of the matrix eq. (A41) together and taking into account that $\sum_m \int dt \ \Delta e_m = 0$, we finally obtain

$$r/J \approx \frac{\sum_m \sum_n (Q^{-1})_{mn} \sigma_n e_n^0}{\sum_m e_m^0} = \langle \sigma_n \tau'_n \rangle, \tag{A42}$$

where $\tau'_n = \sum_m (Q^{-1})_{mn}$ obeys eq. (A31).

We conclude that eq. (A30) is very general; it applies to any system, regardless of how the equilibria between the electrons residing in different traps are settled. All that is required for this formula to be correct is that the excitation pulse is sufficiently weak.

It is noteworthy that the response function $g(t)$ of the conductivity setup does not have to be very fast to obtain the integral

$$\Delta A = \int dt \ [\kappa(t) - \kappa_0]. \tag{A43}$$

in eq. (4). Indeed, due to the basic property of convolution,

$$\Delta A_{\exp} = \int dt \ [\kappa(t) - \kappa_0] \otimes g(t) = \Delta A \ \times \ \int dt \ g(t). \tag{A44}$$

Since the last term in eq. (A44) is unity, by the definition of the response function, $\Delta A_{\exp} = \Delta A$ for any such function. In practice, the latter should be sufficiently fast so that slow reactions neglected in the kinetic analysis given above (for example, charge neutralization and electron scavenging) can be ignored. Another important, albeit obvious, consideration is that the background conductivity signal from the cations (holes) and anions (generated via electron scavenging) does not appear in any expression for $\Delta A$ provided that these species are not photoexcited by the laser pulse, as their concentration and the corresponding contribution to the conductivity signal do not change. The latter requires that the electron concentration is low and the second-order recombination of cations with quasifree electrons during the photoexcitation pulse can be neglected.

**Appendix B. One trap vs. many traps.**

In section IV.A, we assumed that the electron trap is well defined. On the other hand, there are "two-state" theories (see, for example, refs. 28 and 31) which postulate a distribution of localized state energies. [28,31,32,33] Schiller [private communication] gives the following general argument in favor of such a distribution: Assume that a liquid is a canonical ensemble of $n$ identical cells which contain no more than one localized electron. The probability of finding a cell with the energy $E$ is $\Pi(E)$, where the distribution $\Pi(E)$ is given by the familiar expression [L. D. Landau and E. M. Lifshitz: Course of Theoretical Physics, Vol. 5; Statistical Physics, Part 1; 3d edition, chapter 12]

$$\Pi(E) = \frac{1}{\sqrt{2\pi}\sigma} \exp\left[-\frac{(E - \langle E \rangle)^2}{2\sigma^2}\right] \tag{B1}$$

with

$$\sigma^2 = kT^2 C_v, \tag{B2}$$

where $C_v$ is the heat capacity of the medium. [31] This dispersion is ca. 120 meV at 295 K. The total energy of the ensemble is $n\langle E \rangle$. If the liquid contains one quasifree electron with the energy $V_0$, this energy changes to $U_{free} = n\langle E \rangle + V_0$ since the electron interacts with the entire ensemble of the cells. By contrast, a localized electron with the energy $E_{loc} = V_0 - \langle E_t \rangle$ interacts with one cell only and the energy of the ensemble is, therefore, given by $U_{loc} = (n-1)\langle E \rangle + E + E_{loc}$. This localized electron is stable if $U_{loc} < U_{free}$, which is equivalent to $E - \langle E \rangle < \langle E_t \rangle$. From this inequality and eq. (B1), one obtains $P_f \approx erfc(\langle E_t \rangle / \sqrt{2}\sigma)$. Eq. (B1) can also be interpreted differently: the binding energies $E_t = U_{free} - U_{loc}$ are given by the normal distribution $p(E_t)$ centred at $\langle E_t \rangle$ with the dispersion $\delta E_t \approx \sigma$.

Note that the dispersion $\delta E_t$ is comparable with the average binding energy $\langle E_t \rangle$ of the localized electron (Sec. V.A). It might be more appropriate, in the spirit of the fluctuation model, to consider the fluctuations of the *binding potential* $U$ (instead of $E_t$), and use eq. (B2) to estimate the dispersion $\delta U$ of the well *depths*. The latter could indeed be several tens of meV provided that the fluctuations of $U$ are mainly due to the thermal vibrations (breathing modes) of the electron bubble. Such estimates can be readily obtained using the approach used by Parshin and Perverzev [62b] to estimate the width of the *1s-1p* absorption band for electron bubbles in liquid $^4$He. Importantly, the dispersion of trap energies is significant for *liquid* alkanes only. For vitreous alkanes at 77 K, $\delta E_t$ given by eq. (B2) is a few tens of meV's, $\delta U << \langle U \rangle$ and, therefore, $\delta E_t << \langle E_t \rangle$. The experiment of McGrane and Lipsky [44] supports this conclusion: The smearing of the binding energy (which, in the bubble model, equals the threshold energy $E_t$ of the absorption band) would have destroyed the Wigner relation for the absorption cross-section (Sec. V.A) which, experimentally, holds for $(E - E_t)$ as small as a few tens of meV. [44]

In the spherical well model, $E_t = U \cos^2(ka)$ and $U = U_p [ka/\sin(ka)]^2$, [9] where $k$ is the wave vector and $U_p = \hbar^2/2m_e a^2$ is ca. 300 meV for $a \approx 3.5$ Å. Carrying out linear expansion near the mean values one obtains that

$$\frac{\delta E_t}{\delta U} \approx \frac{1 + \sqrt{\langle E_t \rangle / \langle U \rangle}}{1 + \sqrt{U_p / \langle E_t \rangle}}. \tag{B3}$$

Using this formula, one can estimate that for $\langle E_t \rangle \approx 180$ meV and $\delta U \approx \sigma \approx 120$ meV, $\delta E_t \approx 50$ meV, i.e., the dispersion of trapping energies around $\langle E_t \rangle$ is considerable.

From a different perspective, it is common in the physics of disordered solids to assume that density of states for traps in the band tail is exponential,

$$g(E_t) = (N_t / kT_c) \exp(-E_t / kT_c), \tag{B4}$$

where $kT_c$ is the distribution width (typically a few tens of meV) and $N_t$ is the total concentration of band-tail traps [see, for example, G. J. Adriaenssens et al., Phys. Rev. B **51**, 9661 (1995); R. Pandya and E. A. Schiff, Phil. Mag. B **52**, 1075 (1985); X. Chen and C.-Y. Tai, Phys. Rev. B **40**, 9652 (1989); P. Pipoz et al. Phys. Rev. B 55, 10528 (1997) and references therein]. Assuming Fermi-Dirac statistics

$$f(E_t) = \left\{ 1 + \exp\left(\frac{E_t - E_{Ft}}{kT}\right) \right\}^{-1} \tag{B5}$$

for thermal population of these traps, where $E_{Ft}$ is the trapped-electron Fermi level (above which the electron is likely to be emitted to the conduction band), the population $p(E_t)$ of a trap with binding energy $E_t$ is given by the product of $g(E_t)$ and $f(E_t)$. To a good approximation, the Fermi energy $E_{Ft}$ for traps can be replaced by the quasi-Fermi level $E_F$ for free electrons determined from

$$e_f = N_c \exp(-E_F/kT), \tag{B6}$$

where $N_c$ is the concentration of thermally accessible delocalized states above the mobility edge (see the footnote). At $kT << E_F$, $p(E_t)$ has a sharp maximum at $E_t \approx E_F$ which explains why $\langle E_t \rangle$ is close to the activation energy of the electron migration ($\approx E_F$). However, as the temperature increases, $p(E_t)$ becomes wide and bell-shaped, just like the distribution postulated by Schiller and co-workers. [28,31] Thus, both the fluctuation and band tail models suggest considerable variation of binding energies at 295 K.

Our simulations, in which the absorption spectra of trapped electrons calculated using the bubble model for a fixed radius *a* were weighted by the two distributions $p(E_t)$ of binding energies given above (eqs. (B1) and (B2) and eqs. (B4), (B5), and (B6), respectively) suggest that in practice, it is nearly impossible to distinguish between the model with a single trap and the model with multiple traps from the absorption spectra alone. The only spectral region which is sensitive to the distribution is the red edge (6-8 µm) which has not yet been explored. The same calculation suggests that the dispersion of the absorption cross sections around 1 µm (to the blue from the absorption maximum) does not exceed the average cross section, i.e., our estimate of $\bar{\tau}_t = \langle \sigma \tau_t \rangle / \langle \sigma \rangle$ should be of the same order of magnitude as $\langle \tau_t \rangle$.

**Figure captions.**

**Fig. 1S.**

Time-dependent d.c. photoconductivity signals from neat methylcyclohexane at 295 K (the signals are weaker than in other figures shown in this paper since a short-path 2 cm conductivity cell has been used to obtain the traces). The liquid is ionized at $t = 0$ by two photons from a 248 nm laser and then the "trapped electrons" (in fact, impurity anions) are photoexcited by a delayed 532 nm laser pulse at $t = t_L$ (6 ns fwhm, 0.5 J/cm$^2$). The delay time $t_L$ is 0.1, 0.2, 0.4, 0.8 and 1.6 µs, respectively. The signal $\kappa(t)$ obtained in the absence of the 532 nm photoexcitation is plotted to the left; the 532 nm photon induced signal $\Delta\kappa$ is plotted to the right. The decay kinetics of both signals are identical and the amplitude of the $\Delta\kappa$ increases with the delay time of the 532 nm pulse in the same way $\kappa$ decreases as the electron is scavenged by impurity. This behavior indicates that the 532 nm signal is due to the electron photodetachment from an impurity anion generated when the impurity reacts with the electron in methylcyclohexane.

**Fig. 2S.**

(a) The same data for *n*-hexane as in Fig. 2 in the text plotted on the logarithmic time scale. The $\Delta\kappa$ traces (scattered dots; to the right) were fit by a sum of a Gaussian (for the electron spike) and a Gaussian convoluted with an exponential (for the slow component due to the electron photodetachment from an impurity anion). The time constant of this exponential component is the same as that for the $\kappa(t) - \kappa_{ion}$ trace (to the left). The weights of the fast component (the "spike") are juxtaposed on this kinetic trace. (b) The same plot for neat methylcyclohexane at 295 K. The decay of $\kappa(t) - \kappa_{ion}$ is almost perfectly exponential (the experimental data are indicated by a black curve; the exponential fit is given by a faint yellow curve). The vertical arrow in both plots indicates the minimum delay time at which the 1064 nm laser was pulsed. Both saturated hydrocarbons exhibit very similar behavior following the 1064 nm photoexcitation of their ionized solutions.

**Fig. 3S.**

(a) The solid wavy line (scale on the left) is the conductivity signal $\kappa$ from neat *n*-hexane photoionized by 248 nm light (295 K). The dots (scale on the right) are the $\Delta\kappa$ signal due to the 1064 nm photon induced electron detachment from an impurity anion at $t_L \approx 760$ ns (6 ns fwhm, 1.5 J/cm$^2$). The time profile of this signal is identical to that of $\kappa(t) - \kappa_{off}$ except for $t \approx t_L$ where a small spike from photodetrapped electrons is observed. The smooth solid line drawn through the dots is the least squares exponential curve; the extrapolation of this curve to $t \approx t_L$ gives the signal $\Delta\kappa(t_L)$ from the impurity anion. The latter is plotted as a function of 1064 nm photon fluence in part (b) of the plot (empty circles). The solid line in this plot is an exponential curve.

**Fig. 4S.**

Arrhenius plots for the drift mobilities of the electron (empty triangles, to the right; after ref. 4) and molecular ions (to the left) in *n*-hexane obtained using time-of-flight conductivity. The anion mobility $\mu_-$ and cation mobility $\mu_+$ (filled squares and diamonds, respectively) are taken from Table 12.3 and Fig. 12.31 of ref. 51 (at 295 K, we obtained very similar estimates for ion mobilities to those given in that study). The empty diamonds are cation mobilities given from Gee and Freeman, ref. 52 The empty circles indicate the sum mobility $\mu_i = \mu_+ + \mu_-$ of the molecular ions.

**Fig. 5S.**

(a) Radial plot for the probability density $\rho(r)$ of the *1s* ground state of the electron in a spherical potential well for $a$=3.36 Å and $E_t$=180 meV (using formulas from ref. 9). (b) The same as Figs. 5(a) and 5(b) in the text; the molar absorptivity $\varepsilon$ of the electron in the spherical well is plotted as a function of the wavelength.

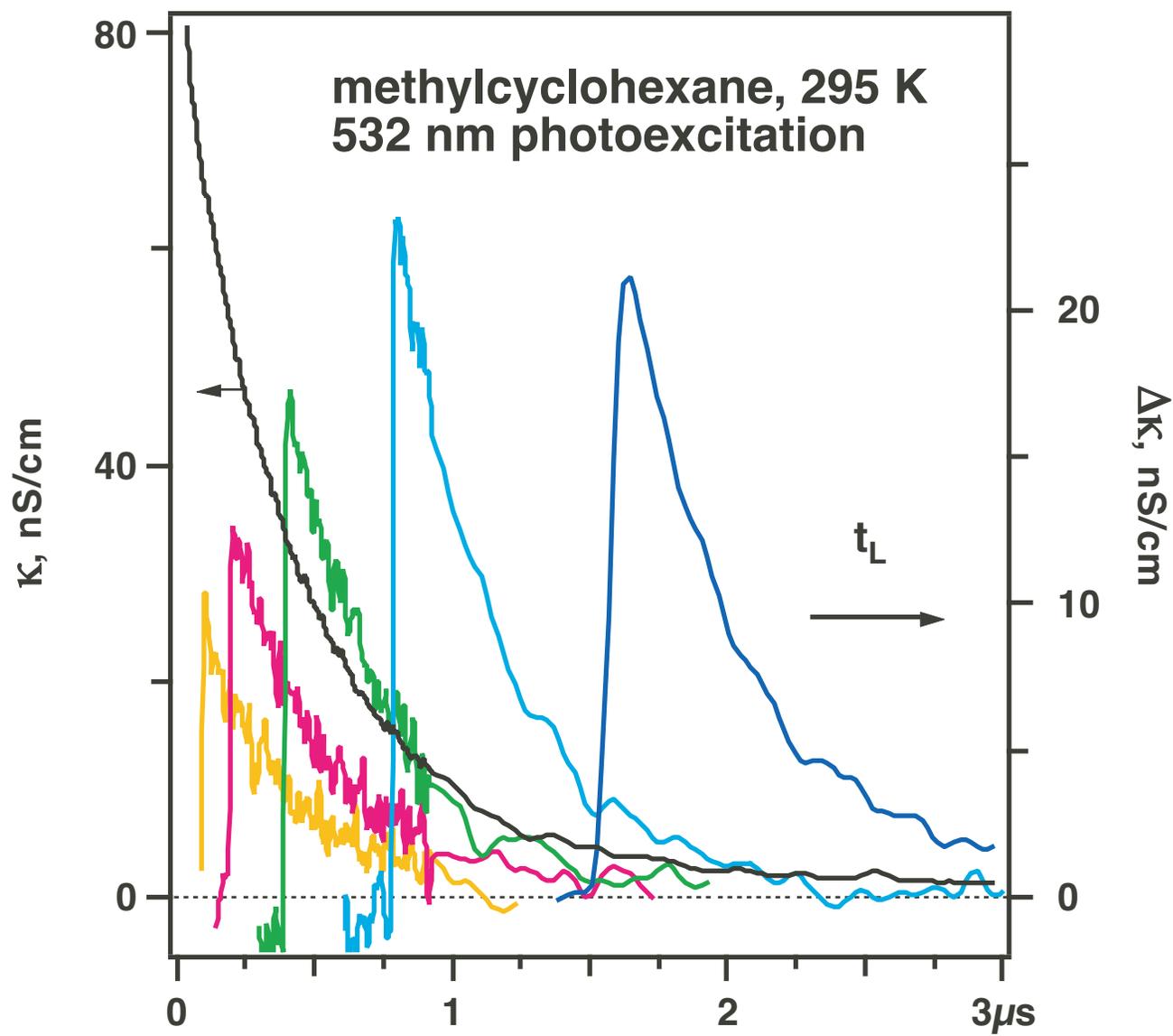

Figure 1S; Shkrob & Sauer

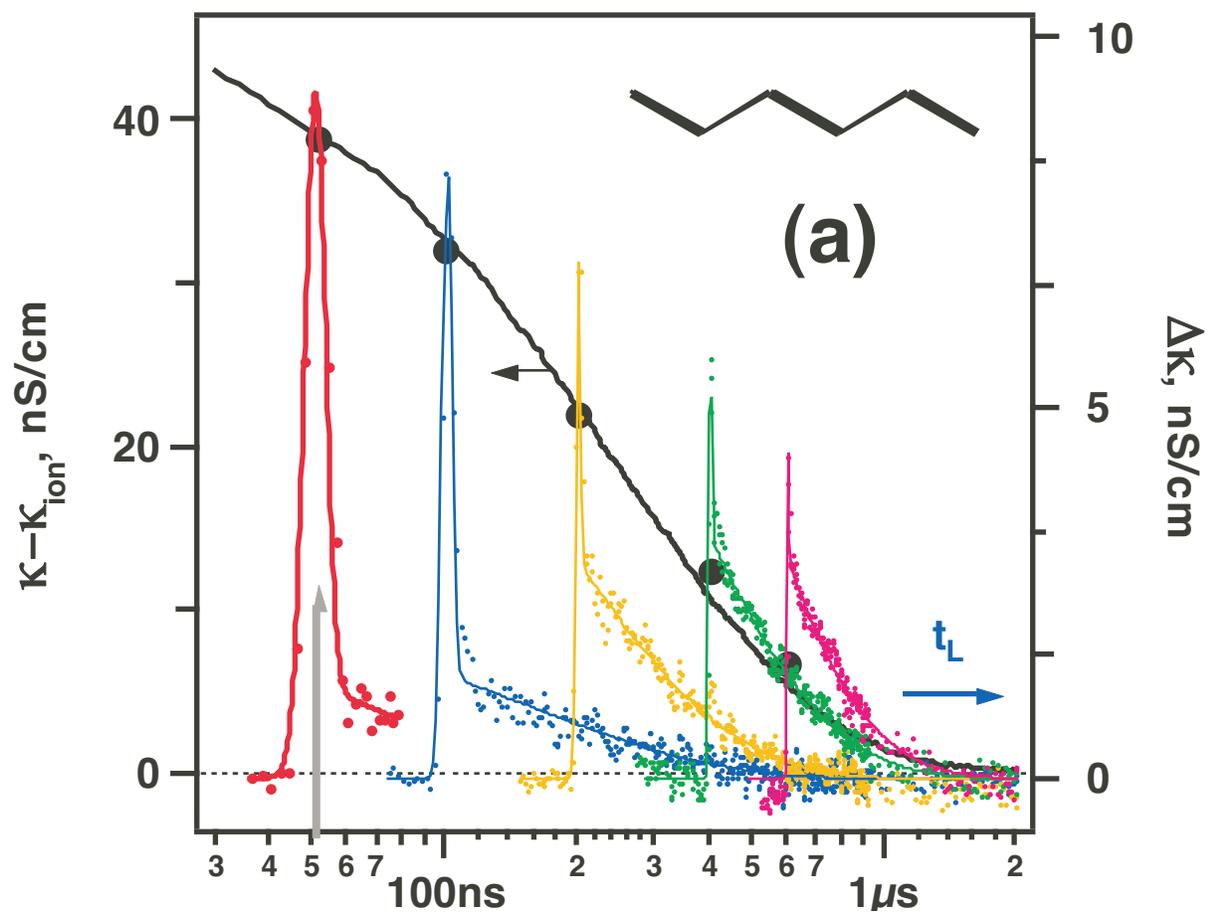
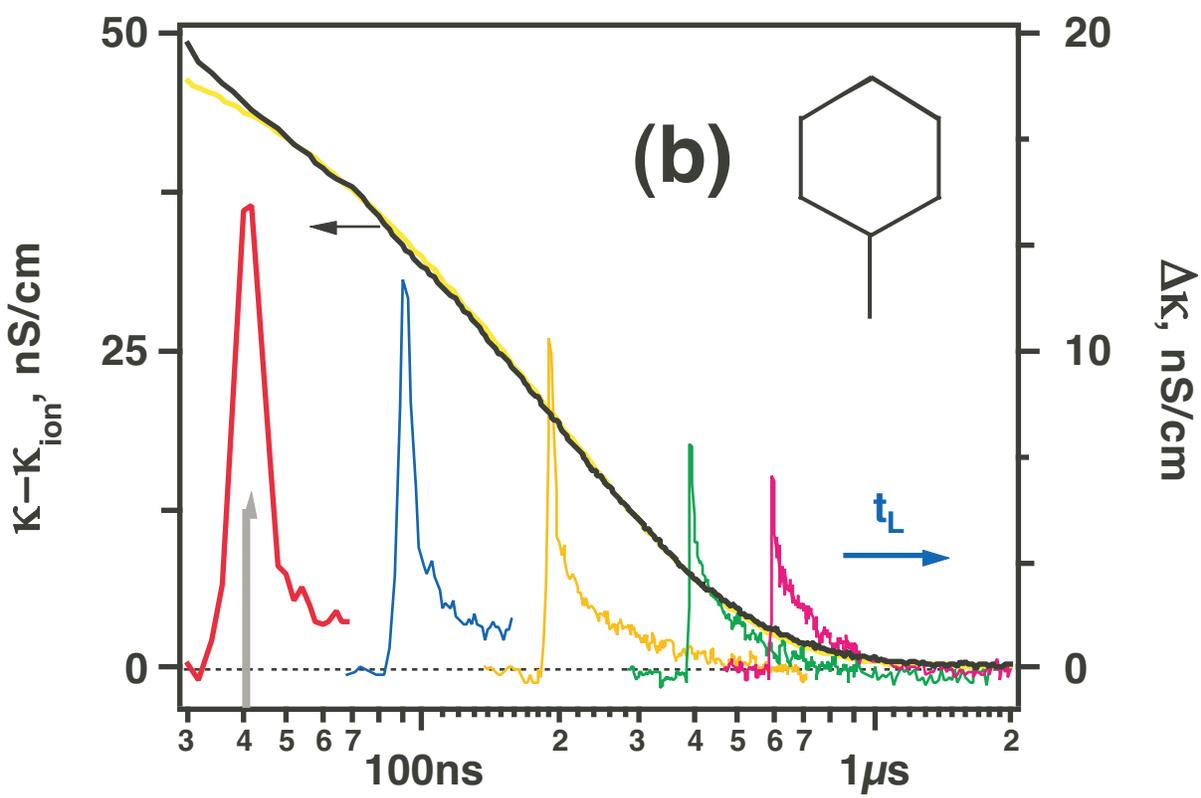

Figure 2S; Shkrob & Sauer

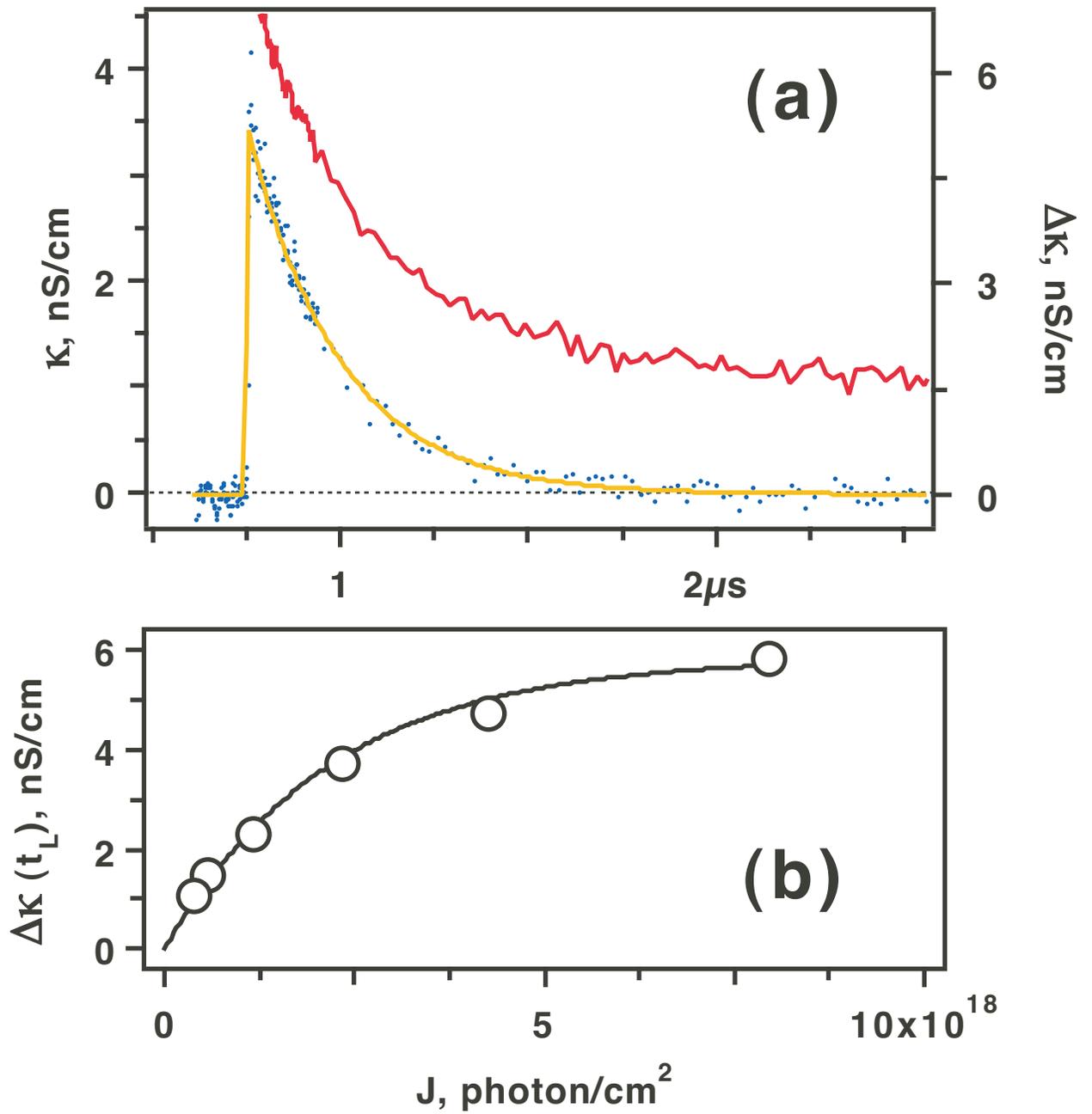

Figure 3S; Shkrob & Sauer

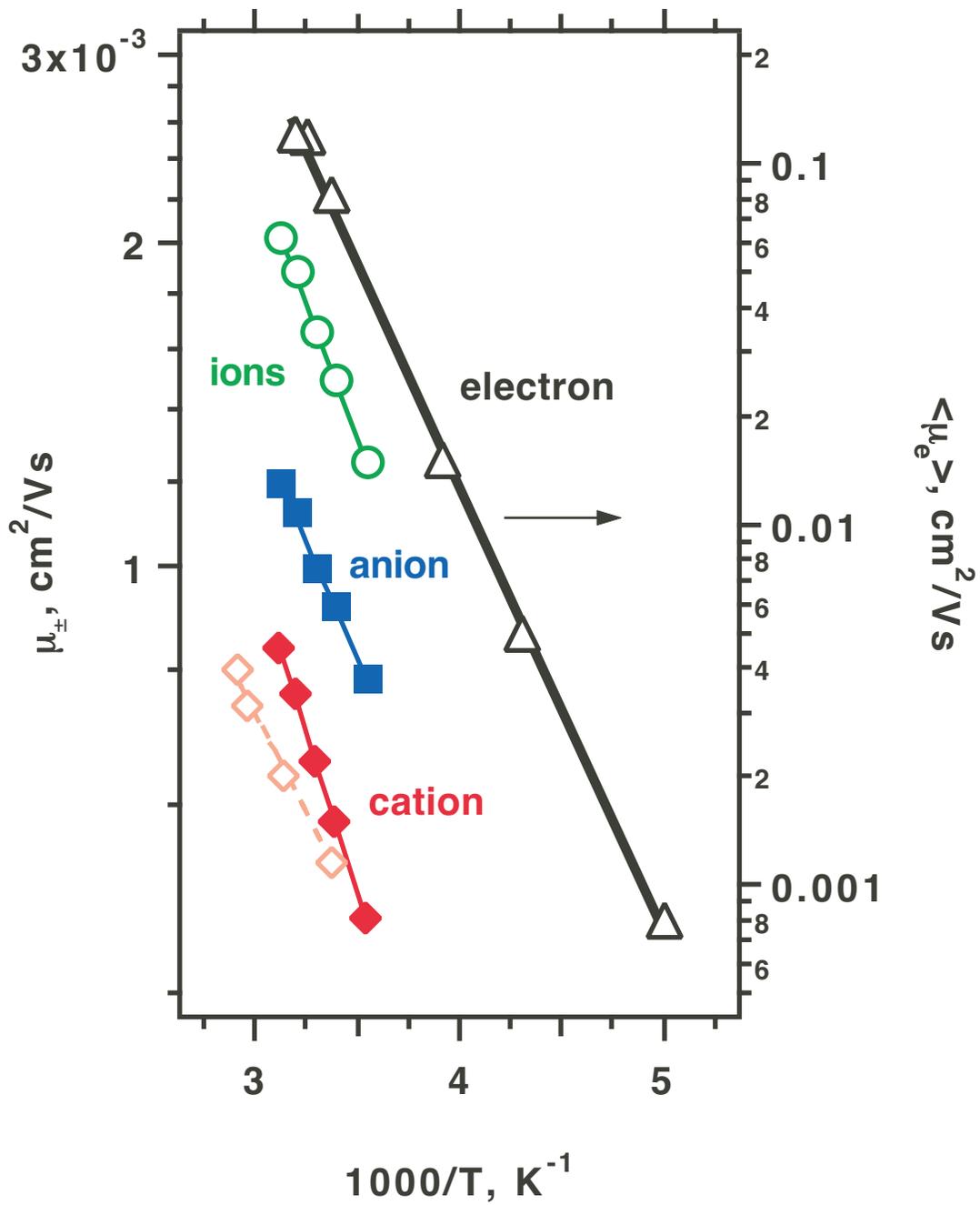

**Figure 4S; Shkrob & Sauer**

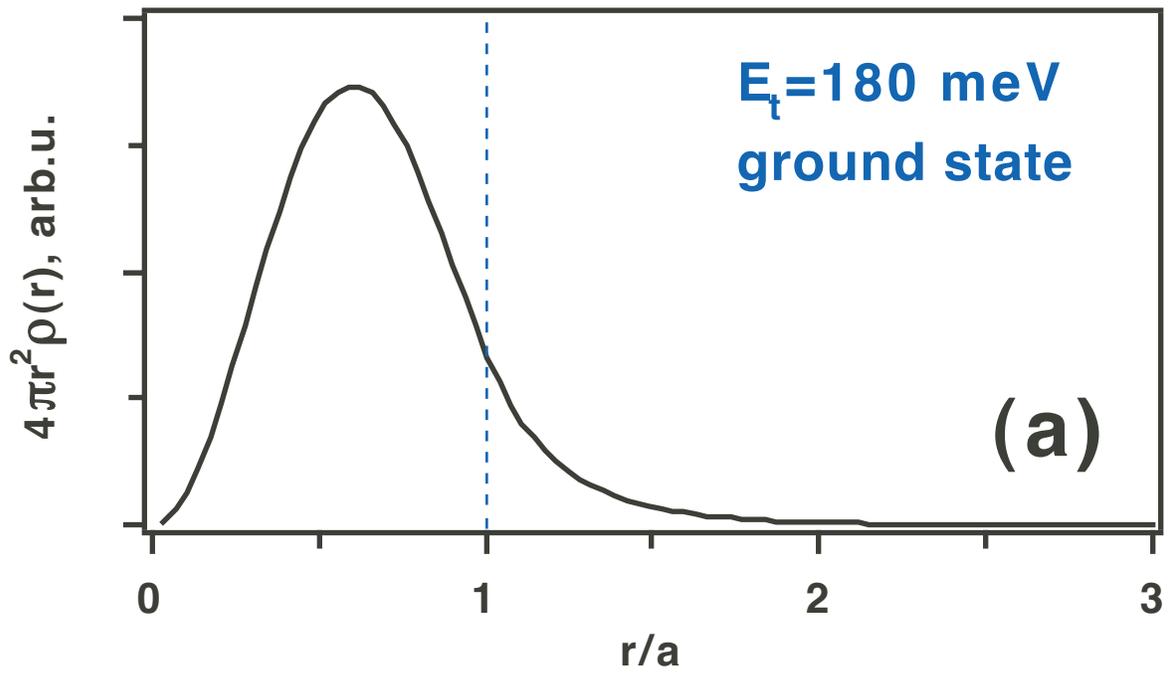

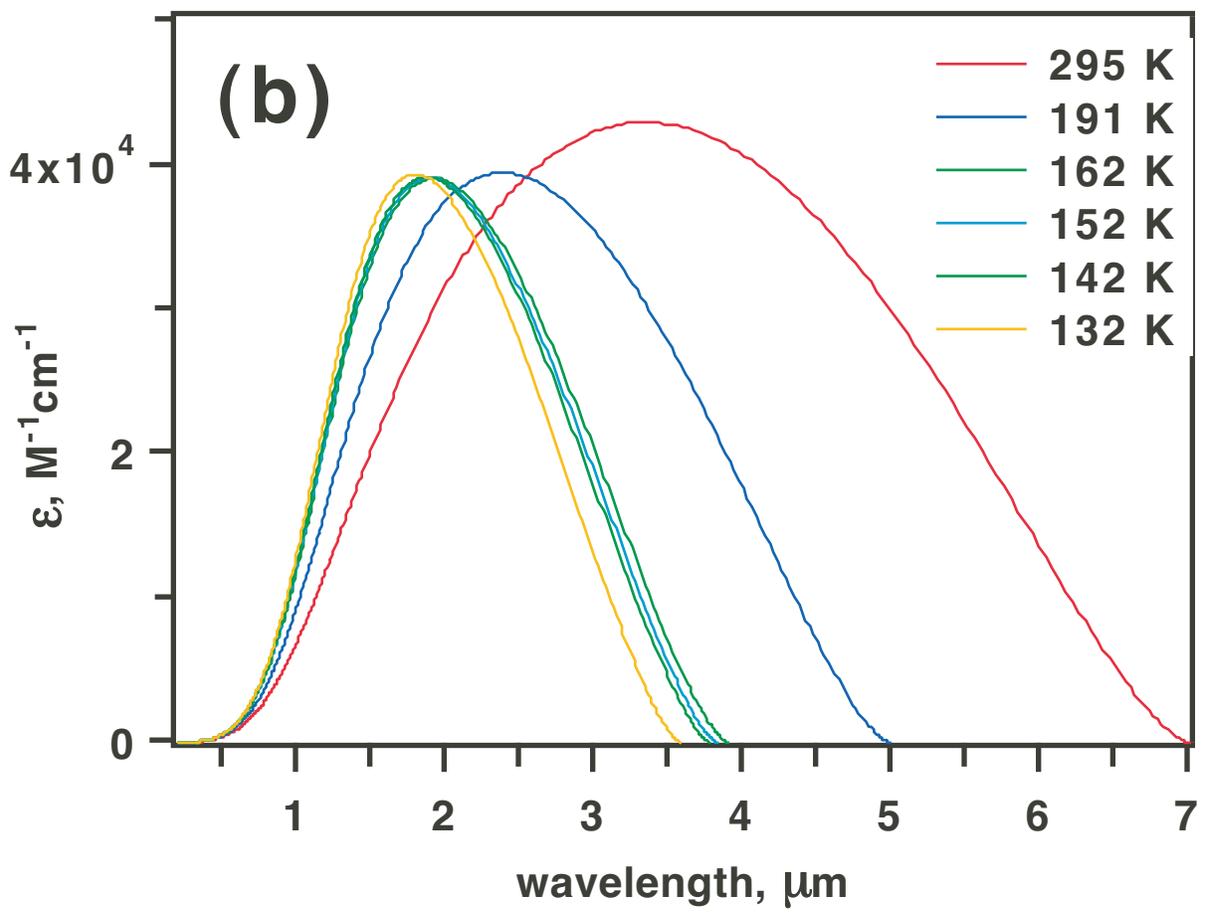

**Figure 5S; Shkrob & Sauer**